\newcommand{\PaperTitle}{Fetch, Read, Reuse: How AI Search Agents Bend the Web’s Rules}

\documentclass[10pt,sigconf,letterpaper,nonacm]{acmart}

\usepackage{makecell}
\usepackage{tikz}
\usepackage{multirow}
\usepackage{tabularx}
\usepackage{rotating}%
\usepackage{booktabs}
\usepackage{array}
\usepackage{xcolor}
\usepackage{colortbl}
\usepackage{siunitx}
\usepackage{listings}
\usepackage[ruled,vlined,linesnumbered]{algorithm2e}

\usepackage[htt]{hyphenat}
\usetikzlibrary{positioning, arrows.meta, fit, calc, backgrounds}

\newcommand{\para}[1]{\vspace{1mm}\noindent\textbf{#1}}

\newcommand{\parai}[1]{\vspace{1mm}\textit{#1}}

\newcolumntype{L}[1]{>{\raggedright\arraybackslash}p{#1}}

\definecolor{badcell}{RGB}{252,210,210}
\definecolor{goodcell}{RGB}{208,238,208}
\definecolor{mixedcell}{RGB}{255,238,190}
\definecolor{nacell}{RGB}{238,238,238}
\definecolor{sechdr}{RGB}{220,228,242}

\newcommand{\origA}{Origin~A}
\newcommand{\origB}{Origin~B}

\newcolumntype{G}[1]{>{\centering\arraybackslash}p{#1}}%
\newcommand{\cellbad}{\cellcolor{badcell}{\boldmath$-$}}
\newcommand{\cellgood}{\cellcolor{goodcell}{\boldmath$+$}}
\newcommand{\cellmixed}{\cellcolor{mixedcell}{\boldmath$\pm$}}

\newcommand{\cmark}{\checkmark}
\newcommand{\xmark}{$\times$}

\newenvironment{smartlist}{%
    \begin{list}{\textbf{$\bullet$}}{%
        \setlength{\topsep}{-\parskip}\addtolength{\topsep}{0pt plus 0.5pt}
        \setlength{\partopsep}{0pt plus 0.5pt}%
        \setlength{\itemsep}{0pt plus 0.5pt}%
        \setlength{\parsep}{0pt plus 0.5pt}%
        \setlength{\leftmargin}{1.3em}%
        \setlength{\labelwidth}{0.7em}%
        \setlength{\labelsep}{0.5em}%
        \setlength{\listparindent}{0pt}%
    }%
}{%
    \end{list}%
}

\definecolor{pubbg}{RGB}{208,224,248}%
\definecolor{vendbg}{RGB}{244,214,236}%
\definecolor{userbg}{RGB}{198,236,242}%
\definecolor{cdnbg}{RGB}{224,214,246}%

\newcommand{\Pub}[2]{& \cellcolor{pubbg}Publisher & #1 & \cellcolor{pubbg}#2 \\}
\newcommand{\Vend}[2]{& \cellcolor{vendbg}Vendor & #1 & \cellcolor{vendbg}#2 \\}
\newcommand{\User}[2]{& \cellcolor{userbg}End user & #1 & \cellcolor{userbg}#2 \\}
\newcommand{\CDN}[2]{& \cellcolor{cdnbg}CDN / infra & #1 & \cellcolor{cdnbg}#2 \\}

\newcommand{\codeurl}{\url{https://github.com/sands-lab/aisearch-probe}}
\newcommand{\dataurl}{\url{https://github.com/sands-lab/aisearch-data}}

\newcommand{\cell}[2]{\shortstack{#1\\[-1pt]{\scriptsize\itshape #2}}}

\newcommand{\cellL}[2]{\shortstack[l]{#1\\[-1pt]{\scriptsize\itshape #2}}}

\renewcommand{\para}[1]{\noindent\textbf{#1}}

\renewcommand{\parai}[1]{\textit{#1}}

\begin{document}

\title{\PaperTitle}

\author{Yixi Chen}
\affiliation{%
  \institution{KAUST}
  \country{\relax}
}
\author{Zafar A. Qazi}
\affiliation{%
  \institution{LUMS}
  \country{\relax}
}

\author{Marco Canini}
\affiliation{%
  \institution{KAUST}
  \country{\relax}
}

\author{Suhaib A. Fahmy}
\affiliation{%
  \institution{KAUST}
  \country{\relax}
}

\begin{abstract}
AI search agents form a new and fast-growing class of web traffic.
Like a browser, they fetch pages on demand for one user; like a transcoding proxy, they rewrite what they fetch; and like a shared cache, they answer other users' questions from the stored copy.
The rules publishers rely on, such as \texttt{robots.txt}, \texttt{Retry-After}, and \texttt{Cache-Control}, were written for more familiar clients, so the agent's vendor decides which of them to follow.
We present a black-box measurement study of how these agents fetch, read, and reuse pages, covering the search APIs of OpenAI, Anthropic, xAI, and Google and the consumer apps of ten services.
We find that OpenAI and xAI do not honor \texttt{robots.txt}, and no service acts on \texttt{Retry-After}.
Services silently read less than the origin served: OpenAI drops some formats, no service reads JavaScript-only content, and xAI reads only the first 0.5\,MB of a page.
OpenAI, Anthropic, and xAI share fetched pages across users and largely ignore \texttt{Cache-Control}, so answers can reflect outdated pages for 30 minutes to a day or more.
Publishers therefore lack effective control: directives are advisory and followed selectively, describe the served page rather than the rewritten copy, and bound how long a copy is reused but not how many answers it serves.
We release our code and data.
\end{abstract}

\maketitle

\section {Introduction}

People increasingly get information from answers written by AI, whether from an assistant such as ChatGPT or from the summaries that search engines now place above their results.
A study of ChatGPT usage found that seeking information, a close substitute for web search, grew from 14\% to 24\% of messages in a single year~\cite{openai2025chatgptusage}, and Google reports strong growth in queries to its AI Mode~\cite{kopack2026alphabet,lang2026alphabet}.
With such an answer on the screen, few people go on to the sources.
In a Pew Research Center study, Google users clicked a link inside a AI summary in only 1\%~\cite{chapekis2025aisummaries}.

The pages behind these answers are still read, but increasingly by the AI instead of the person.
Fetches triggered by questions to AI assistants grew more than 15-fold on Cloudflare's network over 2025~\cite{belson2025yearinreview}, and by mid-2025 such fetchers already made up about a fifth of the AI bot traffic Fastly observed~\cite{fastly2025threatinsights}.
Publishers, meanwhile, receive few visitors in return, as AI platforms request far more pages than they refer visits back~\cite{cloudflare2026radar,tome2025crawltoclick,belson2025crawl}.
A growing share of the web is thus read not by people but by AI assistants that search and fetch pages on their behalf, forming a new class of web traffic. We refer to these as \emph{AI search agents}.

How these agents behave matters because the web runs on a loose contract between servers and clients: a server states its rules, and clients largely follow them.
A site lists in \texttt{robots.txt} what crawlers may read, uses \texttt{Cache-Control} to say how long a copy of a page may be reused, and sends \texttt{Retry-After} to ask clients to back off.
Each rule was written with a familiar kind of client in mind.
A browser fetches a page to show it to one user, a crawler fetches many pages for a search index or a training corpus, and a shared cache stores copies to serve many users.

\begin{table*}[t]
\centering
\caption{Key findings, the parties they affect, and why (our assessment). \autoref{tab:stakeholders-rationale} gives the reasoning for each party.}
\label{tab:stakeholders}
\footnotesize
\definecolor{fetchA}{HTML}{E6EEF8}\definecolor{fetchB}{HTML}{D6E3F3}
\definecolor{readA}{HTML}{EEE9F6}\definecolor{readB}{HTML}{DFD7EE}
\definecolor{reuseA}{HTML}{FAF0DF}\definecolor{reuseB}{HTML}{F3E2C6}
\setlength{\tabcolsep}{4pt}
\renewcommand{\arraystretch}{1.2}
\begin{tabular}{@{}L{1.0cm} L{1.0cm} L{3.9cm} L{2.6cm} L{7.3cm}@{}}
\toprule
\textbf{Stage} & \textbf{Sec.} & \textbf{Finding} & \textbf{Affects} & \textbf{Why it matters} \\
\midrule
\rowcolor{fetchA} Fetch & \autoref{sec:fetcher-id} & \texttt{robots.txt} not honored (OpenAI, xAI)
  & Publisher
  & A publisher's \texttt{robots.txt} opt-out does not stop these fetchers. \\
\rowcolor{fetchB} & \autoref{sec:failure-retries} & \texttt{Retry-After} ignored (all); xAI sends parallel requests
  & Publisher, CDN
  & An origin cannot ask these fetchers to back off, so retries and parallel requests add load to the origin and CDN. \\
\rowcolor{readA} Read & \autoref{sec:content-type-filtering}, \autoref{sec:javascript} & OpenAI drops some formats (e.g. JavaScript not read)
  & Publisher, user
  & Part of what the page served never reaches the answer: the publisher’s content goes unseen, and the user is not told.. \\
\rowcolor{readB} & \autoref{sec:info-access-accross-doc-size} & xAI reads only the first 0.5\,MB
  & Publisher, user, vendor
  & The cap bounds the vendor's cost, but the rest of a long page never reaches the answer, and the user is not told. \\
\rowcolor{reuseA} Reuse & \autoref{sec:content-reuse-and-sharing} & Pages shared across accounts (OpenAI, Anthropic, xAI)
  & Publisher, vendor, user
  & One fetch serves many users: less load and cost, but a publisher paid per fetch is paid once and users may get old copies. \\
\rowcolor{reuseB} & \autoref{sec:cache-directives}, \autoref{sec:retention-and-renewal} & \texttt{Cache-Control} largely ignored (OpenAI, Anthropic, xAI)
  & Publisher, user, CDN
  & Updates and corrections reach answers 30 minutes to a day or more late, and the answer does not show the copy’s age. \\
\rowcolor{reuseA} & \autoref{sec:neg-cache} & Failed fetches are cached
  & Publisher, user
  & A page stays out of answers after its error is fixed, so users miss content the publisher has restored. \\
\bottomrule
\end{tabular}
\end{table*}

Of these clients, crawlers have drawn the most scrutiny.
AI training has made crawling more costly: Wikimedia reports that AI crawlers favor rarely read pages, which miss its regional caches and fall through to its origin datacenters~\cite{wikimedia2025crawlers}, and prior work measures whether crawlers follow publishers' rules~\cite{kim2025scrapers,cui2025odyssey,liu2025somesite}.
Far less is known about AI search agents, which fit none of these roles cleanly. Instead, they combine three roles: they fetch pages on demand for one user (a browser), rewrite what they fetch (a transcoding proxy), and answer other users' questions from the stored copy (a shared cache).
We therefore model an agent as three stages (fetch, read, and reuse) and ask which existing rules govern each:\begin{smartlist}
\item{Fetch.} Like a browser, it fetches a page on demand for one user. Unlike a browser, it hands the page to software rather than to the user, which arguably brings it under \texttt{robots.txt}. Like any client, it should back off when a server sends \texttt{Retry-After}.
\item{Read.} Like a transcoding proxy, it rewrites the page before use, here into a form the model can consume. No HTTP rule governs what it may drop along the way or what it must tell the user.\footnote{The closest is \texttt{Cache-Control: no-transform}~\cite{rfc9111}, which restricts what an intermediary may change, not what the client that consumes the page does with it or tells its user.}
\item{Reuse.} Like a shared cache, it may answer other users' questions from a stored copy, which is what \texttt{Cache-\allowbreak Control} is meant to govern.
\end{smartlist}

An agent only partly fits each role, so its vendor can decide which of that role's rules apply, and no rule yet says what the agent as a whole should do.
Vendors already disagree, starting with \texttt{robots.txt}.
OpenAI states that because \texttt{ChatGPT-User} is initiated by a human and not used for automatic crawling, ``\texttt{robots.txt} rules may not apply''~\cite{openai2026bots}.
Anthropic says its bots respect \texttt{robots.txt}, with no exemption for user-initiated fetches~\cite{anthropic2026crawler}, and xAI's documentation does not say either way~\cite{xai2026websearch}.

The cost of these choices falls on parties with no say in them.
Publishers face bursts of fetches with no reliable way to govern how their pages are accessed~\cite{wikimedia2025crawlers}, and the CDNs and hosting infrastructure behind them absorb the same bursts~\cite{fastly2025threatinsights}.
Once a page is fetched, the publisher also loses control of it: the copy may be rewritten, stored, and used to answer other users, so a publisher that charges per fetch~\cite{teixeira2026payperuse} cannot tell how many answers a single fetch served.
Users cannot tell whether an answer came from an outdated copy or, when a cited answer is wrong, whether the fault lies with the page, the retrieval, or the model.
Citations offer little help, as generative search engines often fail to support their statements with them~\cite{liu2023verifiability,jazwinska2025citation}.
Neither publishers nor users can see what the agent did with the page.

We expose this behavior from the outside through a black-box measurement study of how commercial AI search agents fetch, read, and reuse web pages.
We test the search APIs of OpenAI, Anthropic, xAI, and Google, and the consumer apps of ten services, used through free, logged-in browser accounts: ChatGPT, Claude, Gemini, Grok, DeepSeek, Le Chat, Qwen, Meta AI, Perplexity, and Copilot.
We serve test pages from origins we control and log every request.
Each page carries a unique code, which lets us tie every answer to the request that fetched its source and tell a fresh fetch from a reused copy.
We validate the results on public pages and release our test pages, probe code, and measurement data (\autoref{sec:open-science}).\footnote{Code: \codeurl{}, data: \dataurl{}}

Our findings follow the same three stages; \autoref{tab:stakeholders} summarizes them and the parties each one affects.
\begin{smartlist}
\item{Fetch.} OpenAI and xAI do not honor \texttt{robots.txt} (\autoref{sec:fetcher-id}). No service acts on a \texttt{Retry-After} header, and services otherwise handle failure differently: OpenAI retries server errors, Anthropic gives up after one failure, and xAI sends parallel requests (\autoref{sec:failure-retries}).
\item{Read.} What a service reads can differ from what the origin served. OpenAI drops some formats after fetching them, no service reads content that appears only through JavaScript, and xAI reads only the first 0.5\,MB of a page without telling the user (\autoref{sec:content-type-filtering}, \autoref{sec:javascript}, \autoref{sec:info-access-accross-doc-size}).
\item{Reuse.} OpenAI, Anthropic, and xAI answer one user's question from a page fetched for another and largely ignore \texttt{Cache-Control} (\autoref{sec:content-reuse-and-sharing}, \autoref{sec:cache-directives}). After a page changes, answers can reflect its old version for about 30 minutes to a day or more (\autoref{sec:retention-and-renewal}). Failures are reused too: a page that once failed can stay out of answers after it is fixed (\autoref{sec:neg-cache}).
\end{smartlist}

Taken together, these results point to three gaps in publishers' control over this traffic.
Existing directives are advisory, and vendors follow them selectively.
A directive describes the page the origin served, not the rewritten copy a service keeps.
And \texttt{Cache-Control} bounds how long a copy may be reused, not how many answers it may serve.
We discuss which of these gaps existing mechanisms can close and where new ones are needed (\autoref{sec:http-mechanisms}).

\section{Methodology}
\label{sec:setup}
\begin{figure*}[t]
\centering
\begin{tikzpicture}[
  box/.style={
    draw, rectangle, rounded corners=2pt,
    minimum height=1.25cm, inner sep=4pt,
    align=center, font=\small\sffamily, line width=0.7pt
  },
  cbox/.style={box, minimum width=2.1cm, draw=black!65, fill=blue!5},
  ubox/.style={box, minimum width=2.55cm,
               draw=orange!60!black, fill=orange!5, line width=0.75pt},
  orbox/.style={box, minimum width=2.1cm,
              draw=orange!60!black, fill=orange!5, dashed, line width=0.75pt},
  fbox/.style={draw=orange!60!black, fill=white, rectangle, rounded corners=2pt,
               minimum height=0.6cm, minimum width=1.7cm, inner sep=2pt,
               align=center, font=\footnotesize\sffamily\bfseries, line width=0.6pt},
  cdnb/.style={box, minimum width=1.3cm, minimum height=0.9cm,
               draw=black!40, fill=white, dashed, line width=0.6pt},
  rbox/.style={draw=black!65, rectangle, rounded corners=3pt, line width=0.7pt,
               fill=green!6, minimum height=0.85cm, minimum width=2.5cm,
               font=\small\sffamily\bfseries, align=center},
  arr/.style={-{Stealth[length=5pt,width=3.5pt]}, line width=0.65pt},
  ret/.style={-{Stealth[length=4pt,width=3pt]}, draw=black!40, line width=0.5pt},
  bi/.style={{Stealth[length=5pt,width=3.5pt]}-{Stealth[length=5pt,width=3.5pt]},
             line width=0.65pt},
  ctrlline/.style={
    {Stealth[length=4pt,width=3pt]}-{Stealth[length=4pt,width=3pt]},
    draw=black!40, dashed, line width=0.55pt},
  ann/.style={font=\scriptsize\sffamily, align=center},
]

\node[cbox] (client) {\textbf{Client}};

\node[orbox, right=1.7cm of client] (or)
  {\textbf{OpenRouter}\\[2pt]{\scriptsize optional}};

\node[ubox, minimum width=2.9cm, minimum height=2.2cm, right=0.5cm of or] (vendor) {};
\node[font=\small\sffamily\bfseries] at ([yshift=-0.5cm]vendor.north) {Service};
\node[fbox] (fetcher) at (vendor.center) {Fetcher};
\node[ann] at ([yshift=0.42cm]vendor.south) {OpenAI $\cdot$ xAI\\Anthropic $\cdot$ Google};

\node[cdnb, anchor=west] (cdn) at ($(vendor.east)+(2.2cm,1.15cm)$)
  {\textbf{CDN}\\[2pt]{\scriptsize cache bypass}};
\node[cbox, minimum height=0.9cm, right=0.9cm of cdn] (originA)
  {\textbf{Origin A}\\[1pt]{\scriptsize tagged pages $\cdot$ access log}};

\node[cbox, minimum height=0.9cm, anchor=west] (originB) at (originA.west |- vendor.east)
  {\textbf{Origin B}\\[1pt]{\scriptsize tagged pages $\cdot$ access log}};

\node[cdnb, anchor=west] (pcdn) at ($(vendor.east)+(2.2cm,-2.05cm)$)
  {\textbf{CDN}\\[2pt]{\scriptsize unknown}};
\node[ubox, minimum width=2.1cm, minimum height=0.9cm, right=0.9cm of pcdn] (pori)
  {\textbf{Public site}\\[1pt]{\scriptsize third-party origin $\cdot$ no log}};

\coordinate (xm) at ($(client.center)!0.3!(originA.center)$);
\coordinate (orchpos) at (xm |- originA.north);
\node[rbox, above=0.55cm of orchpos] (orch) {\textbf{Orchestrator}};

\begin{pgfonlayer}{background}
  \node[draw=green!35!black, fill=green!5, rounded corners=4pt,
        densely dashed, line width=0.5pt, inner sep=7pt, fit=(orch)] {};

  \draw[draw=blue!35, fill=blue!4, rounded corners=4pt,
      densely dashed, line width=0.5pt]
  ([xshift=-7pt, yshift= 7pt]client.north west) coordinate (lzone) rectangle
  ([xshift=13pt, yshift=-13pt]client.south east);

  \node[draw=orange!40!black, fill=orange!3, rounded corners=4pt,
        line width=0.5pt, inner sep=10pt, fit=(or)(vendor)] (uvzone) {};

  \node[draw=blue!35, fill=blue!4, rounded corners=4pt,
        densely dashed, line width=0.5pt, inner sep=9pt,
        fit=(cdn)(originA)(originB)] (czone) {};

  \node[draw=orange!40!black, fill=orange!3, rounded corners=4pt,
        line width=0.5pt, inner sep=9pt,
        fit=(pcdn)(pori)] (pzone) {};

  \draw[draw=black!65, fill=blue!5, rounded corners=2pt, line width=0.7pt]
    ([xshift=6pt,yshift=-6pt]client.south west) rectangle
    ([xshift=6pt,yshift=-6pt]client.north east);
  \draw[draw=black!65, fill=blue!5, rounded corners=2pt, line width=0.7pt]
    ([xshift=3pt,yshift=-3pt]client.south west) rectangle
    ([xshift=3pt,yshift=-3pt]client.north east);

  \draw[draw=orange!60!black, fill=orange!5, rounded corners=2pt, line width=0.75pt]
    ([xshift=6pt,yshift=-6pt]vendor.south west) rectangle
    ([xshift=6pt,yshift=-6pt]vendor.north east);
  \draw[draw=orange!60!black, fill=orange!5, rounded corners=2pt, line width=0.75pt]
    ([xshift=3pt,yshift=-3pt]vendor.south west) rectangle
    ([xshift=3pt,yshift=-3pt]vendor.north east);
\end{pgfonlayer}

\node[ann, text=blue!50!black, anchor=north west] at ([xshift=-1pt, yshift=10pt]lzone)
  {\itshape ours};
\node[ann, text=orange!65!black, anchor=south west]
  at ([yshift=0pt]uvzone.north west) {\itshape not ours};
\node[ann, text=blue!50!black, anchor=south west]
  at ([yshift=-2pt]czone.north west) {\itshape controlled tests (our origins)};
\node[ann, text=orange!65!black, anchor=south west]
  at ([yshift=-2pt]pzone.north west) {\itshape public checks (third-party sites)};

\draw[arr] ([yshift=4pt]client.east) -- node[above, ann]{question} ([yshift=4pt]or.west);
\draw[arr] ([yshift=4pt]or.east)     -- ([yshift=4pt]vendor.west);
\draw[ret] ([yshift=-4pt]vendor.west) -- ([yshift=-4pt]or.east);
\draw[ret] ([yshift=-4pt]or.west)     -- node[below, ann]{answer} ([yshift=-4pt]client.east);

\draw[bi] ([yshift=7pt]fetcher.east) -- ++(1.5cm,0) |- (cdn.west);
\draw[bi] (cdn.east) -- (originA.west);
\draw[bi] (fetcher.east) -- node[above, ann]{direct (no CDN)} (originB.west);
\draw[bi] ([yshift=-7pt]fetcher.east) -- ++(1.5cm,0) |- (pcdn.west);
\draw[bi] (pcdn.east) -- (pori.west);

\draw[ctrlline] (orch.west) -|
  node[above, ann, pos=0.25]{dispatch / record}
  (client.north);
\draw[ctrlline] (orch.east) -|
  node[above, ann, pos=0.25]{provision / observe}
  (originA.north);
\draw[ctrlline] (orch.east) -| ([xshift=22pt]originA.east) |- (originB.east);
\end{tikzpicture}
\caption{Measurement testbed.
A measurement client sends a question to a service (left), directly or through OpenRouter.
The service's fetcher requests pages from three kinds of origin (right): two we operate, Origin A behind a CDN with caching bypassed and Origin B served directly, and public sites, which we observe only through the answer.
For our origins, the orchestrator provisions content and failures and reads the access log.}
\label{fig:testbed}
\end{figure*}
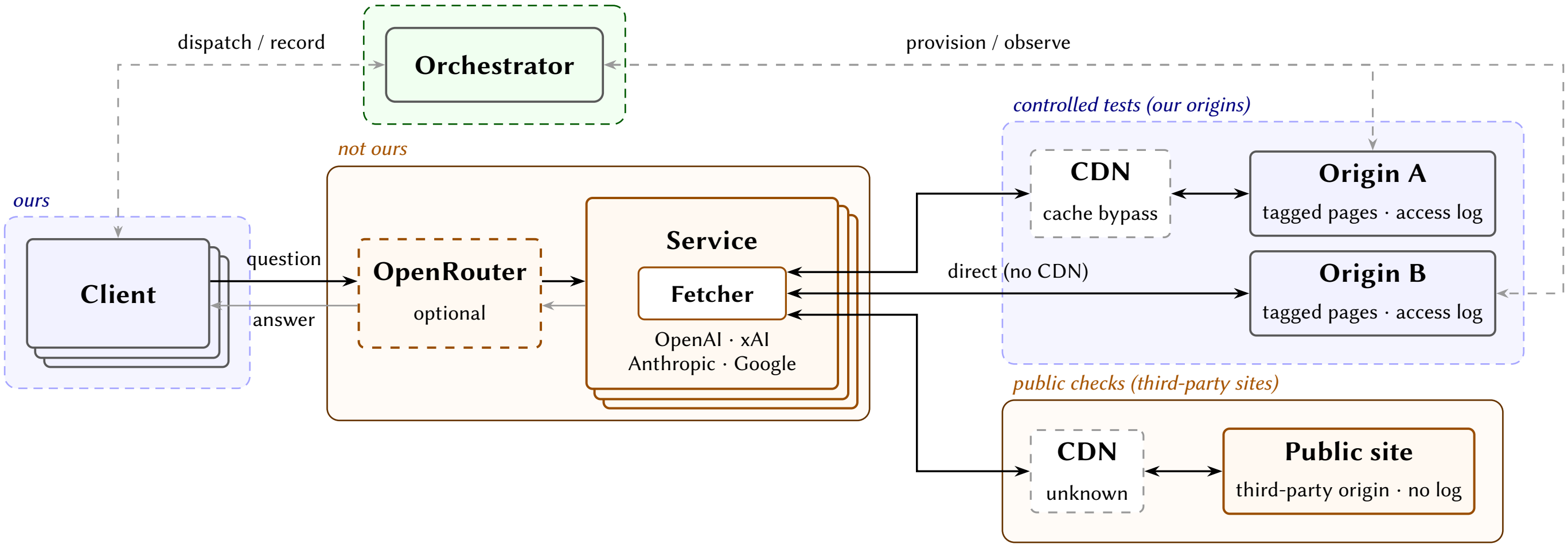

We measure how AI search agents fetch and reuse web pages.
To test various AI services (OpenAI, Anthropic, xAI, or Google), we create a test suite, each consisting of a URL and a question about it; our results are based on the answers to each question.
Specifically, there are two kinds of test:

\begin{smartlist}
  \item \emph{Controlled tests} run against two origins we operate.
  Because we control them, we can change a page, inject failures, set response headers, plant codes in a page, and read every request in the access log, so each behavior can be tied to a controlled change.
  \item \emph{Public checks} repeat selected tests on third-party sites.
  We cannot read their logs or change their pages, so a public check can observe only the answer.
  Of our findings, only two can be checked this way: file-type handling and answer staleness.
\end{smartlist}

The rest of this section follows the path of a test: the services we ask (\autoref{sec:setup-services}), the origins and sites we point them at (\autoref{sec:setup-targets}), the pages and questions we send (\autoref{sec:setup-probes}), and how we run and repeat the tests (\autoref{sec:setup-procedure}).

\subsection{Services Under Test}
\label{sec:setup-services}

We test the AI search services of four vendors: OpenAI, Anthropic, xAI, and Google.
The controlled tests reach them through their APIs; a separate set of tests uses their consumer apps.
We call each tested system a \emph{service}, the HTTP client inside the service to reach origins is its \emph{fetcher}, and the company behind the service is its \emph{vendor}.

\para{API access.}
We query OpenAI's \texttt{gpt-5-mini} and xAI's \texttt{grok-4.3} through OpenRouter, and Anthropic's \texttt{claude-sonnet-5} through Anthropic's own API.
OpenRouter forwards each vendor's own fetcher, so the origin sees the vendor's fetcher and not OpenRouter's (\autoref{app:openrouter}).
An \emph{account} is one API key, at a vendor or at OpenRouter.
Each experiment uses one account per service, except the sharing experiments, which need two: one direct vendor key and one OpenRouter key, with the same model (\autoref{sec:content-reuse-and-sharing}).
For each call, we record the reply, the \emph{input token} charged,\footnote{Tokens are the pieces of text a model reads, and APIs bill by them; input tokens cover our question and the fetched page URL.} any error code, and the fetch tool's own report where the API gives one, such as Google's fetch status or Anthropic's retrieval time.

We use \texttt{gemini-3.1-flash-lite} through Google's free-tier API and report it separately from the other three.
The one exception is the same-caller reuse check on our controlled origins, which used \texttt{gemini-2.5-flash} because \texttt{gemini-3.1-flash-lite} did not give usable answers for these (elaborated in \autoref{app:gemini-origins}).

\para{Consumer apps.}
We also tested the apps of Claude, ChatGPT, Gemini, Grok, DeepSeek, Mistral (Le Chat), Qwen, Meta AI, Perplexity, and Copilot, using free, logged-in browser accounts.
A researcher entered each question into the app and returned the full reply for processing results.
Web search was enabled wherever the app had a switch; apps without one use search on their own.
For the apps, the sharing test used two accounts, one belonging to an author and one to a volunteer (\autoref{sec:content-reuse-and-sharing})%

\subsection{Test Targets}
\label{sec:setup-targets}

\para{Controlled origins.}
We operate two origins, \origA{} (\texttt{probe.\allowbreak attentionhub.org}) and \origB{} (\texttt{agentfetchlab.\allowbreak com}), which serve the same content and differ only in how they deliver it (\autoref{fig:testbed}); we call them the controlled origins.
\origA{} sits behind Cloudflare, a widely used CDN, with caching off so that every request reaches our server.\footnote{Its responses carry \texttt{cf-cache-status: DYNAMIC}.}%
Its backend has no public IP and is reached through a reverse proxy.
However, a CDN may rewrite response headers: our \texttt{ETag} did not reach clients on this origin (\autoref{sec:cache-directives}).
Thus, on \origB{}, we serve it directly by Caddy on a public IP, with no CDN or other intermediary.
Neither origin restricts AI fetchers: both serve the same \texttt{robots.txt}, which allows every crawler on every path.\footnote{Cloudflare does not change the file on \origA{}, and all three vendors' fetches reached our server throughout the study.}%
On \origB{}, every page carries an \texttt{ETag} and a \texttt{Last-Modified} date, and the server answers a conditional request using either with a 304 code.
On \origA{}, Cloudflare removes the \texttt{ETag} but passes \texttt{Last-Modified}, and the 304 code still works.
Both origins serve HTTP/2 and advertise HTTP/3.

One orchestrator drives the controlled origins: it provisions content and failures and reads each origin's access log.
We ran every experiment on \origA{} first, to explore the behavior, and then repeated it unchanged on \origB{}, which has no CDN, to check that the findings do not depend on the CDN.%
The findings agree on both origins, apart from two injected faults that Cloudflare reports differently from Caddy (\autoref{app:fault-behavior}).

\para{Public sites.}
Since we do not control content nor HTTP serving, here we aim to measure what file types AI search handles and whether fresh content is consistently used.
Public tests use three sets of third-party targets.
For file types, we use \emph{public files}: sitemaps, feeds and \texttt{llms.txt} files announced by popular sites.
For staleness, we use two sets.
\emph{Public files} test file types: sitemaps, feeds and \texttt{llms.txt} files announced by popular sites.
The other two sets try to answer the same question: does a service fetch the latest page, or answer from an old copy?

\emph{Changing pages} are homepages of popular sites that change often.
They cover many sites, but we can only tell whether the answer matches the current page or an older one.
\emph{Counter endpoints} are public endpoints that return one value that only rises, such as a blockchain's block height.
They cover fewer sites, but the value in the answer tells us exactly how old the copy is.
How we choose them is in \autoref{app:public-sites}.

\subsection{Probe Design}
\label{sec:setup-probes}

This section describes the controlled tests.
Public checks use third-party pages we cannot change, so none of it applies to them (\autoref{app:public-sites}).

\para{Pages and codes.}
Our default test page is a 2\,KB HTML document generated from a line-based source of frozen prose.
The same source can be rendered as plain text, Markdown, JSON, XML, CSV, PDF or DOCX (\autoref{app:test-documents}).
Each experiment varies one thing about the page, such as its size (up to 2\,MB)%
  , media type or response status, and holds everything else fixed.
To attribute each answer to the page, run, and service that produced it, and to show that the service read the page and did not guess or recall it, we embed unguessable codes in each page, as earlier cache-deception studies plant markers to see where content ends up~\cite{mirheidari2020cached}.
The codes are derived from the page ID and a secret known only to the origin and the measurement client.
Some experiments place five codes at known depths of a longer page, or deliver them through scripts and data blocks (\autoref{fig:known-codes}).
We elaborate on how we construct these pages in \autoref{app:test-documents}.

\para{Questions.}
Each test sends a URL and one question.
Unless an experiment needs a different answer, the question asks for the page's ID and the code at a given depth, quoted exactly (\autoref{app:prompts}).

\para{Attribution.}
Every page has a fresh ID, and its URL carries a service tag.\footnote{
In the format \texttt{https://<origin>/<page-id>.html?<service-tag>}
}
We match the client's questions to origin requests by URL and time, so attribution does not depend on a fetcher's User-Agent or source IP.
The services treat URLs with different tags as different pages (\autoref{app:url-equivalence}), so one page can be asked about under several tags without interference.

\para{Reuse.}
To tell reuse of a page from reuse of an earlier answer, our reuse tests ask successive questions about different codes on the same page.
The later code is absent from earlier answers, so a correct answer with no new origin request shows that the page itself, not the earlier answer, was reused.

\subsection{Procedure}
\label{sec:setup-procedure}

\para{Runs.}
Unless otherwise stated, we test each condition in at least three independent runs, each with fresh page IDs.
We started runs between 15 September and 10 November 2026, mostly between 06:00 and 20:00 UTC, and about a quarter on a weekend.%
The ethical considerations are in \autoref{sec:ethics}.

\para{Timeouts.}
Each API call has a five-minute timeout; a call that exceeds it is recorded as failed.

\para{Where we measure from.}
The measurement client runs on a server in Saudi Arabia, except for our own downloads of public pages, which ran from a server in the United States.

\section{Results}

We organize the results by the three phases of a retrieval, which are also the three actions of the title: \emph{fetch}, how a service requests a page from the origin (\autoref{sec:behavior}); \emph{read}, what it does with the page after fetching it and before answering   (\autoref{sec:content}); and \emph{reuse}, how it keeps and reuses the page across later questions (\autoref{sec:cache}).
Each subsection reports the controlled test first and, where one is possible, a check on public sites (\autoref{sec:setup-targets}).
Two findings have such a check: file-type handling (\autoref{sec:content-type-filtering}) and answer staleness (\autoref{sec:retention-and-renewal}).
The others need the origin's log or control over the page, which public sites do not give us.

\subsection{Fetch: Identity and Failure Handling}
\label{sec:behavior}
Servers set access rules for the clients that fetch from them, and these rules can depend on the client's role.
Here we test how the services meet them: who they say they are, how many requests each question generates, and how they respond to \texttt{robots.txt} and to failures.

\subsubsection{Fetcher Identity and \texttt{robots.txt}}
\label{sec:fetcher-id}
\label{sec:robot-txt-compliance}

Earlier work studies whether crawlers follow \texttt{robots.txt}~\cite{liu2025somesite,kim2025scrapers,cui2025odyssey}.
We ask whether the same holds for the fetches AI search agents make for a user.
A server first has to know who is fetching.
The IETF Crawler Best Practices draft recommends that crawlers identify themselves in the User-Agent header~\cite{illyes2026cbcp}, but the header is self-declared and cannot be verified~\cite{kaiserman2025crawlerspoofing}; Cloudflare’s Web Bot Auth uses signatures instead~\cite{conroy2026agenticweb}.
Prior reports describe Grok's fetcher as impersonating browsers and repeating requests~\cite{stackfox2026grokua,datadome2025spoofing}, so we cannot rely on the declared identity.
Instead, we put a service-specific tag in every URL we send, so every origin request can be attributed to a service (\autoref{sec:setup-probes}).

\begin{table}[t]
\centering
\footnotesize
\setlength{\tabcolsep}{3pt}
\caption{How the API fetchers identify themselves and how many requests they send per question (median).
For Anthropic, 99.9\% is the share of page requests preceded by a \texttt{robots.txt} request within the hour.%
}
\label{tab:fetcher-id}
\begin{tabular}{@{}llrrr@{}}
\toprule
Service & User-Agent & \makecell[r]{Requests\\per question} & \makecell[r]{Source\\IPs} & \makecell[r]{Requests\\\texttt{robots.txt}} \\
\midrule
OpenAI    & \texttt{ChatGPT-User} & 1 & 235     & No \\
Anthropic & \texttt{Claude-User}  & 1 & 47      & Yes (99.9\%) \\
xAI       & macOS browser         & 6 & 7{,}774 & No \\
\bottomrule
\end{tabular}
\end{table}

\para{Identity and request multiplicity.}
OpenAI and Anthropic send vendor-specific User-Agent names, while xAI sends browser User-Agent strings,\footnote{On both origins, xAI's requests used one of two strings around equally, a macOS Safari one and a macOS Chrome one.} so the header does not identify requests from xAI.
xAI's fetcher also sends a median of six requests per question.
In contrast, OpenAI's and Anthropic's send one.
xAI's requests come from far more source addresses, spread over more than a thousand networks (ASNs), while OpenAI's and Anthropic's come from one or two cloud networks.
In a manual check of 100 random xAI requests, only 38 came from hosting providers, so most do not come from data centers (\autoref{app:source-networks}).

This matches the prior reports of Grok's browser impersonation and repeated requests~\cite{stackfox2026grokua, datadome2025spoofing}; \autoref{sec:failure-retries} shows that xAI sends these requests in parallel, whether or not the origin fails.

\para{\texttt{robots.txt}.}
Only Anthropic's fetcher requests \texttt{robots.txt}.
A \texttt{Claude-User} fetcher reads it within a second before its first page request, and it reuses the rules it retrieved from \texttt{robots.txt}   between page fetches (\autoref{tab:fetcher-id}).
We observed no such request from \texttt{ChatGPT-User}, and none attributable to xAI's browser-like fetcher.\footnote{We identify the vendor by a tag in the test page's URL, which we cannot add to \texttt{/robots.txt}, because the fetcher requests it on its own; so xAI's browser-like User-Agent strings leave no way to attribute such a request.
}
This agrees with what the vendors state~\cite{openai2026bots,anthropic2026crawler}.
How long \texttt{Claude-User} keeps the rules before reading them again is in \autoref{app:robots-reuse}; whether the consumer apps request \texttt{robots.txt} is in \autoref{app:robots-apps}.

\subsubsection{Retries and Backoff Under Failure}
\label{sec:failure-retries}
The status code of a failed response tells a web client what to do next.
After a temporary failure (\texttt{500}, \texttt{503}, \texttt{429}) a client may retry, and a \texttt{Retry-After} header can say how long to wait first, although a client is not required to comply~\cite{rfc9110,rfc6585}.
After a permanent one (\texttt{403}, \texttt{404}) a retry is pointless, and a redirect loop should be cut off.
We serve each of these from our origin and record the requests that follow, to see whether a fetcher retries or stops, whether it waits as told, and how many requests one question sends.
\autoref{tab:fault-behavior} gives the counts for a \texttt{500} (Internal Server Error), a \texttt{403} (Forbidden) or \texttt{404} (Not Found), and a redirect loop; the \texttt{Retry-After} tests and the other faults are in \autoref{app:fault-behavior}.

\begin{table}[t]
\centering
\small
\caption{  Requests that reached our origin per question; 1 means a single request, with no retry.
\emph{Retry}: a later request after a response; \emph{parallel}: concurrent requests, so not all are retries.
For the redirect loop, the number of redirects followed.
The full table is in \autoref{app:fault-behavior}.}
\label{tab:fault-behavior}
\begin{tabular}{@{}lccc@{}}
\toprule
Origin response & OpenAI & Anthropic & xAI \\
\midrule
\cellL{500 Internal Server Error}{temporary: a retry is allowed}
  & \cell{3--8}{retry} & \cell{1}{no retry} & \cell{5--7}{parallel} \\
\cellL{403 Forbidden, 404 Not Found}{permanent: a retry is pointless}
  & \cell{1}{no retry} & \cell{1}{no retry} & \cell{4--14}{parallel} \\
\cellL{Redirect loop}{should be cut off}
  & \cell{1}{\phantom{x}} & \cell{9}{\phantom{x}} & \cell{57--77}{parallel} \\
\bottomrule
\end{tabular}
\end{table}

\para{No service waits for \texttt{Retry-After}.}
A server sends \texttt{Retry-After} to tell a client how long to wait before trying again.
We attached a ten-second \texttt{Retry-After} to a \texttt{503} (Service Unavailable) and to a \texttt{429} (Too Many Requests).
However, no service waited the ten seconds and then retried: each either did not retry or retried without respecting the delay..
Namely, OpenAI retried the \texttt{503} before the ten seconds were up, and did not retry the \texttt{429}; Anthropic did not retry either.
xAI sent its usual overlapping requests, so no retry stands out.\footnote{We confirmed with \texttt{curl} that \texttt{Retry-After} arrived through Cloudflare to an outside client.
We cannot verify what each vendor's fetcher received.}

\para{xAI sends overlapping requests.}
xAI made four to seven requests even for a \texttt{404} or an empty successful page, and 12--28 when the response was cut off (\autoref{tab:fault-behavior}, \autoref{tab:fault-behavior-full}).
The requests overlap in time, so they are parallel fetches rather than retries.

\para{Anthropic barely retries; OpenAI retries only server errors.}
Anthropic sent only one request in almost every case, so a single failure ended the question.
OpenAI sent three to eight requests after a \texttt{500}, but only one after a \texttt{403} or \texttt{404} (more cases in \autoref{app:fault-behavior}).
Whether a failure is remembered across questions is tested in \autoref{sec:neg-cache}.

\para{Redirect loops.}
Given an endless redirect, OpenAI followed one redirect, Anthropic nine, and xAI issued 57--77 parallel hops.

\para{A failed fetch costs the caller nearly as much as a successful one.}
For the same 2\,KB page, a failed fetch uses about as many input tokens as a successful one, for all three vendors (\autoref{fig:fault-token-cost}, \autoref{app:fault-tokens}).
What accounts for the cost of a failed fetch is not visible from outside the service.

\subsection{Read: Content Handling and Cost}
\label{sec:content}

A fetch that reaches our origin does not mean the service uses the page in its answer.
We test which formats, script-delivered text, and page sizes survive the fetch, and what each costs in input tokens.

\subsubsection{Document Formats}
\label{sec:content-type-filtering}
A service may fetch a document without using its contents in an answer.
We test which document formats the services can read: each file contains codes, and we check whether the answer repeats them.
Each trial uses a new test file in one of eight formats: HTML, plain text, Markdown, JSON, XML, CSV, PDF, or DOCX with five codes (\autoref{app:test-documents}).
We record whether the service requested the file, which codes appear in its answer, and whether it says it cannot read the format.

Changing the format also changes the file's bytes, in both content and length.

To isolate the \texttt{Content-Type} header we serve one byte-identical 2\,KB body under five different values and ask for its codes.

\autoref{tab:content-type} shows which documents each service used in its answer.
Every service fetched every document, and the results are the same on both origins.

\begin{table}[t]
\centering
\small
\setlength{\tabcolsep}{4pt}
\caption{Which documents do AI fetchers use?
Every fetcher retrieved every file format from our origin, but not every format was used in the answer.
\cmark: the answer returned all five requested page codes; \xmark: the service reported an unsupported content type.  }
\label{tab:content-type}
\begin{tabular}{@{}>{\raggedright\arraybackslash}m{1.65in}%
  >{\centering\arraybackslash}m{0.5in}%
  >{\centering\arraybackslash}m{0.55in}%
  >{\centering\arraybackslash}m{0.3in}@{}}
\toprule
 & OpenAI & Anthropic & xAI \\
\midrule
\multicolumn{4}{@{}l}{\emph{(a) One source, eight formats}} \\
HTML, plain text, JSON, PDF & \cmark & \cmark & \cmark \\
Markdown, XML, CSV, DOCX & \xmark & \cmark & \cmark \\
\midrule
\multicolumn{4}{@{}l}{\emph{(b) One body, five \texttt{Content-Type}s}} \\
\texttt{text/html}, \texttt{text/plain} & \cmark & \cmark & \cmark \\
\texttt{text/markdown}, \texttt{application/xml}, \texttt{text/csv} & \xmark & \cmark & \cmark \\
\bottomrule
\end{tabular}

\end{table}

\para{OpenAI rejects four of eight formats.}
OpenAI answered with all five requested values from HTML, plain-text, JSON, and PDF documents.
For Markdown, XML, CSV, and DOCX, it instead reported an unsupported content type.
Markdown is a notable case: document-ingestion tools such as MarkItDown~\cite{microsoft2024markitdown} and Docling~\cite{auer2024docling} deliberately convert source documents to Markdown for LLM use.
Anthropic and xAI answered from all eight formats.
OpenAI’s fetcher still downloaded all four rejected formats, so the rejection happens after the fetch, not before it.

\para{OpenAI decides by the \texttt{Content-Type} header, not the content.}
The format test changes both the body content and the header, so it cannot tell which one decides whether OpenAI reads the document.
We therefore serve one 2\,KB body, the same byte for byte, under five \texttt{Content-Type} header values.
OpenAI answered under \texttt{text/\allowbreak html} and \texttt{text/\allowbreak plain}, but reported an unsupported content type under \texttt{text/\allowbreak markdown}, \texttt{application/\allowbreak xml}, and \texttt{text/\allowbreak csv}.
Anthropic and xAI answered under all five.
With the bytes fixed, the header alone decided whether OpenAI used the page.

\para{The request headers accept every format.}
Anthropic sent \texttt{Accept: */*}; OpenAI and xAI sent browser-style lists with a wildcard fallback.
OpenAI's list included \texttt{application/xml}, although it rejected the XML response.

\para{On public sites.}
Public files show the same pattern: OpenAI and Google refuse some XML files, while Anthropic and xAI read them (\autoref{app:public-xml}).
OpenAI refused 98\% of XML files, served as \texttt{application/xml} or \texttt{text/xml}, but read 95\% of plain-text files from the same kind of sites; its answers quote the fetcher's ``Unsupported content-type'' error.
Google refused every \texttt{application/xml} file but returned content for most \texttt{text/xml} files, and its error is generic;\footnote{Every refusal carried \texttt{URL\_RETRIEVAL\_STATUS\_ERROR} and every successful read \texttt{URL\_RETRIEVAL\_STATUS\_SUCCESS}.
The error value names no cause; it does rule out Google's separate statuses for unsafe or paywalled content.
Google is included here only: it is excluded from the controlled tests (\autoref{sec:setup}).} because both are XML, the \texttt{Content-Type} header is the likely trigger, though Google does not say so.
Anthropic refused plain text (19\%) about as often as XML (14--19\%) and blamed the site's domain, not the format.
xAI rarely refused, but returned 47\% of the \texttt{text/xml} files flattened into one line.

\subsubsection{JavaScript-Dependent Content}
\label{sec:javascript}

Many sites insert their main text with client-side JavaScript, so a fetcher that does not run scripts sees a different page from a browser user.
Vercel reports that most AI crawlers do not render JavaScript~\cite{zecchini2024aicrawler}; we test the user-triggered fetchers directly, with test pages that hide some codes behind a script.
We describe the test pages and runs in \autoref{app:test-documents}.

\begin{table}[t]
\centering
\footnotesize
\setlength{\tabcolsep}{4pt}
\caption{Do AI fetchers see text that JavaScript puts on a page?
Only the plain-HTML code ever appeared in answers, for all three services.}
\label{tab:js-variants}
\begin{tabular}{@{}L{1.8cm} L{2.7cm} L{1.8cm} L{1.7cm}@{}}
\toprule
\textbf{Version} & \textbf{Where the other codes sit} & \textbf{Readable in page source?} & \textbf{In answer?} \\
\midrule
Plain HTML (baseline) & HTML body text & Yes & Yes \\
Inline script & Scrambled in an inline script & No & No \\
External script & Scrambled in a \texttt{.js} file & No & No \\
Data block & JSON block, copied in by a script & Yes & \textbf{No} \\
JSON-LD & JSON-LD block, never displayed & Yes & \textbf{No} \\
\bottomrule
\end{tabular}
\end{table}

\autoref{tab:js-variants} lists the five versions.
The plain-HTML version is our baseline; for this, all three services returned all five codes in every sitting, so the page itself is readable.

First, do the fetchers run scripts?
In two versions of the page, the codes are scrambled inside a script (\autoref{tab:js-variants}), so they can be read only by running the page's decoder.

No service returned them.
No service requested the imported \texttt{.js} file, including xAI, which sent five to seven page requests per question with browser User-Agents (\autoref{sec:fetcher-id}).
We saw no sign that any fetcher runs page scripts.

Second, do they at least read text inside a script tag?
In the data-block and JSON-LD versions, the codes are in the HTML as plain text, yet no service returned them.
This suggests that the fetchers remove the content of \texttt{<script>} tags before the page reaches the model, as HTML-to-text conversion commonly does; we infer this from the answers and cannot observe the pipeline (\autoref{sec:limitations}).
It also differs from Vercel's remark that JSON data in the initial HTML response may still be indexed~\cite{zecchini2024aicrawler}.
Anthropic's fetch tool with dynamic filtering (\autoref{sec:post-op-cost-amp}), in which the model writes a program to analyze the fetched page, did not help either.

Text that a site supplies only through JavaScript or in a script block, such as JSON-LD, is therefore invisible to all three fetchers.
About 40\% of the popular-site homepages we checked embed JSON-LD, so this content is likely missing from many answers, although we did not measure this over public pages.

\subsubsection{Document Size}
\label{sec:info-access-accross-doc-size}
\label{sec:token-scaling}
A fetch of a long page does not show that the whole page informs the answer, or what it costs.
We serve pages from 4\,KB to 2\,MB with codes at known depths (\autoref{sec:setup-probes}), and record which codes the answer reports and the input tokens the API charges.
Note that no fetcher sent a \texttt{Range} header, although our origin supports byte ranges: every request asked for the whole page.

\begin{figure*}[t]
  \centering
  \includegraphics{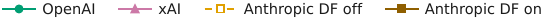}\\[2pt]
  \includegraphics{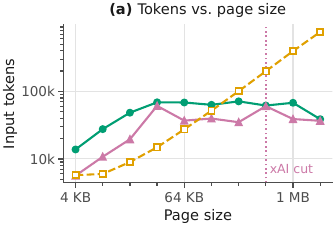}\hfill
  \includegraphics{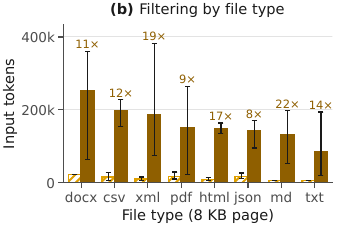}\hfill
  \includegraphics{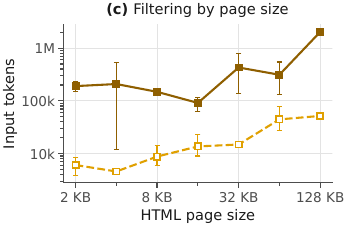}
  \caption{Input tokens (page content given to the model).
  (a) Tokens vs.\ page size; the dotted line is where xAI stops reading (${\sim}$500\,KB).
  (b)--(c) Anthropic with dynamic filtering (DF) on or off, by file type for an 8\,KB page and by HTML page size.
  Points and bars are means of up to 3 runs, whiskers show min and max.
  The label above each DF-on bar is its mean divided by the DF-off mean.}
  \label{fig:df-panels}
\end{figure*}

\para{xAI reads only the first half megabyte.}
OpenAI and Anthropic returned every code at every size up to 2\,MB.
xAI did the same through 512\,KB, but on the 1\,MB and 2\,MB pages it returned only the codes within the first half megabyte, in every trial on both origins.
Both origins sent the whole page in every one of these requests, so the limit applies to what xAI reads, not to what it downloads.
Because xAI sends several requests per question (\autoref{sec:fetcher-id}), a 2\,MB page is served several times for an answer that uses only its first half megabyte.

\para{Input grows with page size only for Anthropic.}
Anthropic's reported input token count kept growing roughly with page size, to about 760\,K tokens at 2\,MB with no plateau.\footnote{We did not set Anthropic's \texttt{max\_content\_tokens} limit in these runs~\cite{anthropic2025webfetch}.}
OpenAI's and xAI's count rose on small pages, then stayed within a broad band up to 2\,MB (\autoref{fig:df-panels}a).
The two bounded services differ in what they deliver: OpenAI returned all five codes at a bounded cost, whereas xAI's bounded input coincides with the loss of content beyond about 0.5\,MB.

At the small end the order reverses: on a 4\,KB page OpenAI reported more than twice the input of the others, because most of its input is not the page (about nine times our estimate of the page's own tokens, against four for Anthropic and xAI) but search and tool context~\cite{openai2026pricing}.
This is an end-to-end difference, not a statement about how OpenAI reads the page.\footnote{We estimate page-content tokens from Anthropic's 2\,MB run.
Ratios for OpenAI and xAI are approximate because their tokenizers differ.}
These runs used the basic fetch tool; \autoref{sec:post-op-cost-amp} shows how much more the direct path with dynamic filtering costs.

\subsubsection{Dynamic Filtering and Input Cost}
\label{sec:post-op-cost-amp}

A retrieval tool may process a document before the model answers, which the origin log cannot show.
We compare two versions of Anthropic's fetch tool on the same pages, model, and question: the basic version, and a newer one with dynamic filtering, which can execute code to select fetched content~\cite{anthropic2025webfetch}.\footnote{The versions are \texttt{web\_fetch\_20250910} (basic) and \texttt{web\_fetch\_20260209} (dynamic filtering, direct API only).}
We compare API-reported input tokens, tool traces, and answers.

\para{Dynamic filtering increased input on these pages.}
Dynamic filtering aims to reduce input by selecting relevant content from a fetched document.
On our 8\,KB pages, however, the version with filtering reported roughly 8--22 times as many input tokens as the basic version across all eight formats (\autoref{fig:df-panels}b).
Enabling dynamic filtering also makes input tokens vary widely across repeated runs, where they were stable without it.

The HTML comparison shows that the increase persisted at larger page sizes (\autoref{fig:df-panels}c); one 128\,KB call reported about 2 million input tokens.\footnote{This single call cost approximately \$4, so we did not test larger pages; the median public homepage we measured is 158\,KB.}

The filtering calls ran code (their traces contain \texttt{code\_execution} blocks, which the basic calls never did), yet gained nothing: both versions answered the question about the deepest code, and filtering used more input tokens.
Filtering is meant to discard irrelevant content, and our question needs codes from across the whole page, so there was little to discard.
We did not test pages where most of the content is irrelevant to the question, where filtering may save tokens.

\subsection{Reuse: Caching and Sharing}
\label{sec:cache}

We ask whether information fetched for one query is reused later: by whom, for how long, and is this despite what the origin says.
Each test has three steps.
First, a warm-up question makes the service fetch a page we serve.
Then, after a gap, the origin keeps the page, changes it, or makes it fail.
Finally, a repeat question asks for a code that only a new fetch could return, and the origin log shows whether one happened (\autoref{app:reuse-design}).

\subsubsection{Reuse and Sharing Across Callers}
\label{sec:content-reuse-and-sharing}
Does a page fetched for one caller serve another?
After a warm-up question fetches a page, we ask the repeat question from a different caller: a different model, account, or app login.
The repeat question asks for a different code on the same page, so only the page's content, not a replay of the first answer, can answer it.
\autoref{tab:sharing-scope} shows, for each change of caller, whether the service reused the page.
The API and app trials used separate pages, so they do not test reuse between the two.

\begin{table*}[t]
\centering
\small
\setlength{\tabcolsep}{5pt}
\caption{Is a later question answered from a page that an earlier question made the service fetch, with no new request to our origin?
Yes: in every trial (3--6 per cell); No: in none; otherwise the share of trials. ---: not tested (most apps do not let us choose the model, and we did not vary the apps' location).
The ``later caller'' is the one asking the second question.
Gemini API: \texttt{gemini-2.5-flash} (\autoref{app:gemini-origins}).}
\label{tab:sharing-scope}
\begin{tabular}{@{}l cccc @{\hspace{1.6em}} cccc@{}}
\toprule
 & \multicolumn{4}{c}{API} & \multicolumn{4}{c}{Consumer app} \\
\cmidrule(lr){2-5}\cmidrule(l){6-9}
Later caller is\dots & OpenAI & Anthropic & xAI & Gemini & ChatGPT & Claude & Grok & Gemini \\
\midrule
the same account (new query or chat)
  & Yes & Yes & Yes & No & Yes & Yes & 1 of 3 & Yes \\
the same account, a different model
  & Yes & Yes & Yes & --- & --- & --- & --- & --- \\
a different customer
  & Yes & Yes & Yes & --- & Yes & Yes & 2 of 3 & No \\
a different customer, called from the US
  & Yes & Yes & Yes & --- & --- & --- & --- & --- \\
\bottomrule
\end{tabular}
\end{table*}

\para{Reuse across API callers.}
OpenAI, Anthropic, and xAI shared page content across models and accounts in every completed trial (\autoref{tab:sharing-scope}).\footnote{The account comparison paired a direct vendor account with an OpenRouter account using the same model.}
Gemini is the exception: it fetched the page again on every question, even from the same caller minutes later, so it had no copy to share. This matches Google's documentation, which says that its tool reads a page from an index cache when the page is indexed and fetches it live otherwise~\cite{google2026urlcontext}; our pages were new.

\para{Consumer apps show a narrower and less consistent pattern.}
ChatGPT and Claude reused the page in a new chat and shared it across separate logins in all three sittings.
Grok did so inconsistently: it reused the page in one of three sittings and shared it across logins in two, and fetched again otherwise.
The full eight-app comparison appears in \autoref{app:app-sharing}.

\subsubsection{Cache-Control and Conditional Requests}
\label{sec:cache-directives}

HTTP origins can specify when a response may be reused through \texttt{Cache-Control} and \texttt{Expires}~\cite{rfc9111}.
Validators such as \texttt{Last-Modified} let a client check whether a retained response has changed~\cite{rfc9110}.
We test whether these signals affect how often a service reuses a page or requests it from our origin.
We serve the same 2\,KB page under each of 19 header settings and ask about it again after 2.5 to 80 minutes (\autoref{app:cache-settings}).
\autoref{tab:anthropic-nostore} summarizes the outcome.

\begin{table}[t]
\centering
\scriptsize
\setlength{\tabcolsep}{2.5pt}
\caption{Was a page answered from a stored copy, with no origin request, when re-asked up to 80\,min later?
Columns give the page's \texttt{Cache-Control} setting.
OpenAI and xAI behaved identically in every run and for all 19 settings (\autoref{tab:cache-directives}).
OpenAI fetched the page again after about 30\,min under every setting, including none, so this is not a response to the header.
OR = via OpenRouter, D = direct.}
\label{tab:anthropic-nostore}
\begin{tabular}{@{}llcccc@{}}
\toprule
& & \multicolumn{4}{c}{Reused?} \\
\cmidrule(l){3-6}
Run & Route & \texttt{no-store} & \texttt{private} & \texttt{no-cache} & None \\
\midrule
\multicolumn{6}{@{}l}{\emph{Anthropic}} \\
7 Oct                & OR & Yes & Yes & Yes & Yes \\
15 Oct               & D  & Yes & Yes & Yes & Yes \\
17--21 Oct, 6 runs   & D (5), OR (1) & No  & Yes & Yes & Yes \\
\midrule
\multicolumn{2}{@{}l}{\emph{OpenAI}} & Yes & Yes & Yes & Yes \\
\multicolumn{2}{@{}l}{\emph{xAI}}    & Yes & Yes & Yes & Yes \\
\bottomrule
\end{tabular}
\end{table}

\para{The header did not change reuse, except for Anthropic's \texttt{no-store}.}
OpenAI answered from a stored copy for about 30\,min whatever the header said, then fetched the page again.
xAI never fetched a page again during our test, even for pages marked \texttt{no-store} or \texttt{no-cache}.
Anthropic also ignored the header until 15 October.
From 17 October on, in every run and on both routes, it fetched \texttt{no-store} pages again on every question, but still reused pages marked \texttt{private} or \texttt{no-cache}, or with no directive (\autoref{tab:anthropic-nostore}).

\para{No API fetcher revalidated.}
On both origins we observed no \texttt{If-Modified-Since} or \texttt{If-None-Match} request from the three API fetchers: whenever they asked again, they asked for the whole page.\footnote{Among the consumer apps, only Mistral sent a conditional request, one \texttt{If-Modified-Since}, and our origin answered it with \texttt{304 Not Modified}.}

\para{On public pages, Anthropic's copies outlive the header.}
For pages that set a limit on reuse, 76\% of its answers came from a copy older than the header allows.
Even for \texttt{no-store} pages only 42\% followed it, and the rest were answered from copies a day old or older.\footnote{One page that sets \texttt{no-store} was answered from a copy almost six days old, and one that sets \texttt{max-age=60} from a copy a day old.}
Pages marked \texttt{private}, which a cache shared between users should not store, still got an old copy 81\% of the time, and Anthropic shares stored copies between callers (\autoref{sec:content-reuse-and-sharing}).

\subsubsection{Retention and Renewal}
\label{sec:retention-and-renewal}

Without conditional validation (\autoref{sec:cache-directives}), a service may keep answering from an older copy after its source changes.
We first measure when a correction becomes visible in origin requests and answers.
We then compare repeatedly queried pages with pages left untouched to test whether use postpones an origin revisit.

\begin{figure*}[t]
  \centering
  \includegraphics{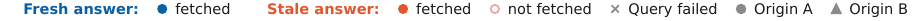}\\[2pt]
  \includegraphics{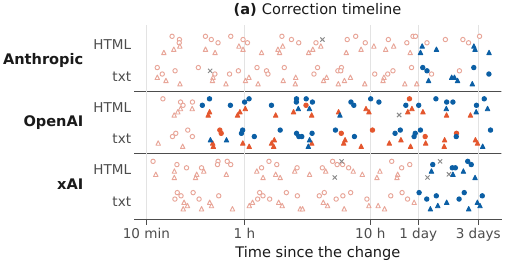}\hfill
  \includegraphics{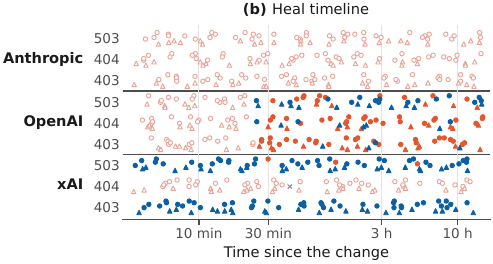}
  \caption{How long after a change a service keeps answering from the old state. Each mark is one follow-up question, placed by the time since (a) a page update or (b) the fix of an error page.
  Stale means the answer still shows the old page (a) or the error (b); fetched means our origin received a request for that question.
  Time axes are logarithmic.}
  \label{fig:timelines}
\end{figure*}

\para{Visibility of page updates.}
We ask about a fresh HTML or plain-text page, update it at the same URL, and ask again after a delay.
OpenAI gave updated answers from about 30 minutes on, while Anthropic and xAI gave outdated answers for a day or more (\autoref{fig:timelines}a).
Notably, even after OpenAI fetched the updated page, about 60\% of its follow-up answers were still outdated; we look into why in \autoref{app:old-after-fetch}.
A fetch in the origin log therefore does not mean the answer uses the page.

\para{Questions did not renew copies.}
Asking about a page several times did not keep its copy longer: OpenAI and xAI downloaded it again at the same final check as a page left alone (\autoref{app:renewal-test}).

\para{On public pages, Anthropic answered from copies that were hours or days old.}
To see whether the reuse we measured on our own pages also happens on real sites, we asked about 90 homepages that change often (elaborated in \autoref{app:public-staleness}).
Anthropic's fetch tool reports when it fetched each page, so we can read the age of its copy directly.\footnote{The reply carries a \texttt{retrieved\_at} time and the exact retrieved content for each fetched page, and a failed fetch returns an error code such as \texttt{url\_not\_allowed} or \texttt{url\_not\_accessible}.
The other three services return no fetch time in our runs (OpenAI and xAI through OpenRouter, Google through its API).}
65\% of Anthropic's answers to the normal URL came from a copy more than an hour old, and 7\% from one more than a week old.
The answers agree: Anthropic answered from outdated copies ($p<0.001$).

\para{Public endpoints agree with our own origins.}

Judging whether a page's text is stale is noisy, since a service may quote a different part or paraphrase it.
A public endpoint whose value only increases has none of this noise: the answer is a number, so how far it lags the live value is measured directly, and we do not need to build a different question for each page.
We asked each service for the current value of such an endpoint, one that changes every few seconds to minutes, every 45 minutes for about three days.
We compared each answer with our reference log, which records the endpoint's value every five minutes.

\begin{figure}[t]
  \centering
  \includegraphics[width=\columnwidth]{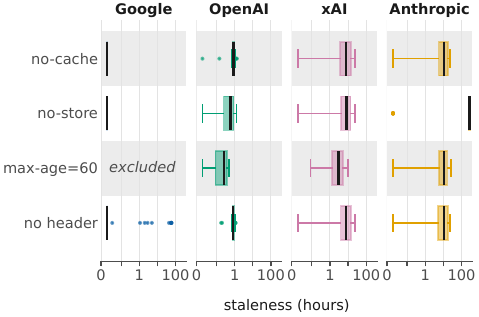}
  \caption{How far behind the live value each service's answer is, on one public endpoint per \texttt{Cache-Control} setting: Hacker News (\texttt{no-cache}), NIST beacon (\texttt{no-store}), USGS quakes (\texttt{max-age=60}), and Ethereum height (no header).
  Google's USGS answers are excluded because it mis-transcribed the value.}
  \label{fig:endpoint-staleness}
\end{figure}

\autoref{fig:endpoint-staleness} shows one representative endpoint for each header setting (\autoref{app:endpoint-staleness} shows all eight).
Google was almost always current.
OpenAI's answers were typically about an hour behind the value in our reference log.

xAI's were about eight hours behind and Anthropic's about eleven, in line with the copies lasting up to about a day on our origins (\autoref{fig:timelines}).

The header made no visible difference, as on our own origins: endpoints marked \texttt{no-store}, \texttt{no-cache}, or \texttt{max-age=60} were no fresher than those with no header.
Notably, we also checked how Anthropic handles \texttt{no-store} on the NIST beacon, and the result only partly agrees with our origins.
Its answers were either current or from a very old copy, with nothing in between; \autoref{app:endpoint-staleness} looks into why.

\subsubsection{Heal from Negative Caching}
\label{sec:neg-cache}

Negative caching is established in DNS: resolvers can retain a failed lookup and answer later queries without asking the authoritative server~\cite{rfc2308}.
We ask whether services do the same with a failed page fetch, after the origin has recovered.
Each trial starts with a fresh URL that returns an error (\texttt{503}, \texttt{404}, or \texttt{403}) throughout the first query; once that query ends, we switch the URL to a marked \texttt{200} page and ask again after a delay.
A service that remembers the failure answers with an error and sends no request; one that does not fetches the page and returns the marker.
A page that was available from the start, asked at the same age, shows whether the service reuses ordinary copies at that delay.

\para{Failures are remembered, but differently by each service.}
Anthropic did not fetch a recovered page again for any of the three errors in the whole window, up to about 11 hours, and its answers kept saying the page was broken.
OpenAI sent no request for about the first 25 minutes; after that it fetched the recovered page again, but 70\% of such trials its answer still quoted the original error.
xAI read the fix promptly for \texttt{503} and \texttt{403}, but for \texttt{404} behaved like Anthropic, with no new request at any delay (\autoref{fig:timelines}b).

\section{Discussion}

\subsection{HTTP Mechanisms in the AI Era}
\label{sec:http-mechanisms}
Our findings show that AI search agents do not follow the assumptions that HTTP makes of browsers and crawlers.
We discuss which existing mechanisms still apply and what new ones the retrieval stack may need.

\para{Control signals work only when a vendor chooses to follow them.}
Cache directives remain the right way for a publisher to say how its pages may be reused, and one vendor already follows part of them: from mid-October 2026, Anthropic honors \texttt{no-store} by fetching such pages again on every question (\autoref{sec:cache-directives}).
Within a single role, such as reuse, an existing directive still applies, and following it takes a policy change by the vendor, not a new protocol.
But compliance is partial.
Anthropic follows \texttt{no-store} on our pages but not consistently on a public endpoint (\autoref{app:endpoint-staleness}), and it still reuses pages marked \texttt{no-cache}; OpenAI and xAI follow none of the directives.
The other signals we tested are no better: OpenAI and xAI ignored \texttt{Retry-After}, and only Anthropic requested \texttt{robots.txt} (\autoref{sec:failure-retries}, \autoref{sec:fetcher-id}).
For \texttt{robots.txt}, the standard defines rules for crawlers, which it calls automated clients, and says nothing about fetches that a user triggers~\cite{rfc9309}.
Nothing in these specifications requires a vendor to follow the advisory signal, so a publisher gives up control over how its pages are fetched and reused to the vendors.

\para{Directives describe the origin's page, not the service's copy.}
HTTP's reuse rules assume that a cache holds the object the origin served: \texttt{ETag} and \texttt{Last-Modified} check whether it changed, and \texttt{Cache-Control} says how long it stays valid.
Services process pages before the model reads them: none read text that a page supplies only through scripts (\autoref{sec:javascript}), and Anthropic's dynamic filtering lets the model run code on the fetched page (\autoref{sec:post-op-cost-amp}).
A \texttt{max-age=600} on a news article governs only the original HTML, not the text a service extracts from it, and no fetcher sent a conditional request (\autoref{sec:cache-directives}).
We cannot see what a service stores, so whether a processed copy is what it reuses remains open.

\para{Publishers cannot tell who is fetching.}
An origin identifies a client by its address and its User-Agent, and both are self-declared.
OpenAI and Anthropic name their fetchers, while xAI's documentation does not~\cite{xai2026websearch} and its requests carry browser-like User-Agents from many source addresses across over a thousand networks (\autoref{sec:fetcher-id}, \autoref{app:source-networks}).
An operator cannot single such a fetcher out by address or identity, and log studies that attribute traffic by declared User-Agent~\cite{kim2025scrapers} would miss it.
Emerging IETF work proposes to bind an agent's identity to a cryptographic signature instead of an address~\cite{meunier2026webbotauth}, which would let an origin apply a policy per agent.
Any mechanism that enforces control above needs this first.

\para{What new mechanisms may be needed.}
Two gaps remain that no existing HTTP mechanism covers.

\textit{A request cannot say how much the answer depends on it.}
AI search agents multiply the requests an origin receives, since one user question can trigger many fetches~\cite{sayyid2026sustainable}.
Some of these requests are exploratory, collecting context by following links, while others fetch the page the answer uses.

To the origin they look the same, so when it is overloaded it cannot serve the needed requests first and slow the rest.
A lightweight header that marks a request as essential or exploratory would allow this.

\textit{HTTP caches whole pages, but retrieval reuses pieces of text.}
An HTTP cache stores a response under its URL and treats it as stale as a whole.
A retrieval system splits pages into chunks and reuses the chunks that are relevant to a question~\cite{jin2026ragcache}, which may come from pages on several origins.
Such a system may answer from a stored chunk without requesting the page, and one changed sentence makes the whole URL stale under HTTP.
How freshness rules should apply to chunks, when one answer mixes chunks from several origins, is an open problem.

\subsection{Vendors See the Fetch, Users and Publishers Do Not}
Vendors see what a fetch retrieved, when, and at what cost.
Users and publishers see little of it, and cannot respond to what they cannot see.

\para{Users cannot see what a fetch costs.}
Input cost depends on the service, not only on the page (\autoref{sec:info-access-accross-doc-size}).
OpenAI's stays within a band up to 2\,MB, xAI bounds it by reading only the first 0.5\,MB and dropping the rest without saying so, and Anthropic's grows with the page, to about 760\,K tokens for a 2\,MB page.
Anthropic offers a size limit~\cite{anthropic2025webfetch}, but a user cannot predict in advance what size is proper for their question.
A failed fetch costs too: for a 2\,KB page, a call whose fetch failed used about as many input tokens as one whose fetch succeeded (\autoref{app:fault-tokens}), so it adds to the bill without adding anything to the answer.

\para{Users cannot see how old an answer is.}
Nothing in an answer shows whether it comes from a live fetch or a stored copy.
An update reached answers after 30 minutes (OpenAI) to a day or more (Anthropic, xAI) (\autoref{sec:retention-and-renewal}), and OpenAI sometimes answered from the old version although it had fetched the new one (\autoref{app:old-after-fetch}).
Only Anthropic's API reports the retrieval time and lets a developer ask for a fresh fetch~\cite{anthropic2025webfetch}; we found neither for OpenAI or xAI, and none of the services shows this to the end user.
From outside, we observe only how long a copy lasts, not whether a fixed time limit or eviction ends it; either way, the answer carries no sign of its age.

\para{Publishers cannot see how their page is used.}
A fetch shows the publisher one request, not how many answers it will serve: a stored copy is shared across accounts (\autoref{sec:content-reuse-and-sharing}), and each later use leaves no request at all.
The origin's log therefore counts fetches and not uses, so anything that depends on counting, such as pay-per-fetch, sees only part of the use~\cite{teixeira2026payperuse}.

\para{Publishers do not know what format an AI agent wants.}
There is no agreed format for web content that is friendly to AI agents.
An agent has two options.
If it wants the human view, it should run scripts and browse interactively.
If it wants the content, it should read the file directly, in a form like Markdown.
The services do neither fully.
Some retrieval stacks browse a page like a person~\cite{openai2025gptoss}, yet no service runs JavaScript (\autoref{sec:javascript}), and OpenAI refuses Markdown (\autoref{sec:content-type-filtering}), the format commonly recommended for LLM input~\cite{microsoft2024markitdown,auer2024docling}.
A publisher therefore cannot tell which form to optimize for.

\subsection{Limitations}
\label{sec:limitations}

\paragraph{Cost-constrained parameter space.}
Our experiments incurred approximately \$400 in API costs, which constrained the parameter space we could explore.
In particular, object sizes beyond \SI{128}{\kilo\byte}, where the token costs are highest, were tested only sparsely.
Real-world HTML pages frequently exceed this threshold, so our cost estimates are conservative for the tail of the size distribution.

\paragraph{Two controlled origins.}
Some experiments need a controlled origin: planting codes, setting arbitrary \texttt{Cache-Control} directives, and varying response size are not possible on third-party sites.
We use two, one behind a CDN with caching bypassed and one served directly, and their results agree, but both are our own new domains.
Agents may apply different policies to established sites, depending on domain reputation, location, or hosting provider, which two origins cannot exercise.
We complement the controlled measurements with observations on public pages and endpoints.

\paragraph{Public-page validation.}
On third-party pages we see only the answer, not the process behind it: there is no request log and no extracted text.
A stale answer therefore shows that an old version was used, but not where it came from.
The fetcher may also receive a different edition of a page than our polling server does (by region, language, or bot rules), or an older copy from the site's own cache.
Services differ in how they turn a page into text and in whether they quote at all, so we score each answer by outcome and report read rates per service.
The sites are popular pages that allow all four fetchers, and a stale answer is a lower bound on reuse.
We therefore use these pages to validate the controlled results, not to estimate rates.

\paragraph{Vendor coverage and black-box view.}
We study three commercial API vendors (Anthropic, OpenAI, xAI), plus Google on public pages and several consumer apps in the appendix; other vendors may behave differently.
Behavior also changes over time: Anthropic's mid-study shift to fetching \texttt{no-store} pages again on our origins shows that findings can become outdated within the measurement period, so our results are a snapshot.
We see the request log at the origin and the API reply at the client, but not what happens in between.
The dynamic-filtering result is inferred from token counts and tool traces, internal steps such as chunking, storing, and deduplication cannot be observed, and our explanation for old answers after a new fetch is a guess (\autoref{app:old-after-fetch}).
For the same reason, we cannot say why Anthropic's fetch tool reused an old copy of the NIST beacon, which sends \texttt{no-store}, but fetched our own \texttt{no-store} pages again (\autoref{app:endpoint-staleness}).
We also measure how long a service keeps a page and whom it shares it with, but not how often reuse happens, which depends on how popular pages are among users (for web pages, popularity is classically skewed~\cite{breslau1999zipf}) and is visible only to the vendors.

\paragraph{Content we did not test.}
Our pages are plain text, and a question asks for a code that a plain string match can find, so the check does not depend on how well the model understands the page.
Images and video would need richer pages, and checking an answer about an image or a video would depend on whether the model can read it, which adds a variance of its own.
We also did not test how services handle ads, which is a natural next direction.
It would mean serving ads to the fetchers, and since services keep stored copies and share them across accounts, those ads could reach other users' answers.

\section{Related Work}

\paragraph{Bot and crawler measurement.}
Kim et al.\ find that scrapers comply less as \texttt{robots.txt} directives get stricter, and that AI search crawlers rarely check the file at all~\cite{kim2025scrapers}.
Cui et al.\ study how publishers write \texttt{robots.txt} rules for LLM bots and report cases where a bot did not follow them~\cite{cui2025odyssey}.
Sayyid et al.\ audit what ChatGPT and Claude can do with a page and the traffic this creates: one question can trigger 10--40 fetches, and fetched pages are reused for about 10--15 minutes from the same address~\cite{sayyid2026sustainable}.
Unlike studies of periodic crawlers and the controls publishers have against them~\cite{kim2025scrapers,cui2025odyssey,liu2025somesite}, or of what answer engines can do with a page~\cite{sayyid2026sustainable}, we study the on-demand fetch itself, beyond whether it honors \texttt{robots.txt}.
We measure it at the origin server, in request logs: what a fetcher requests, how long it keeps and shares what it fetched, how it handles failures, and what each fetch costs.

\paragraph{AI traffic impact on web infrastructure.}
Zhang et al.\ inject controlled AI workloads into a Varnish deployment and find that cache miss rates double when AI traffic reaches 25\% of requests and triple at 100\%~\cite{zhang2025webcache}.
The Wikimedia Foundation reports that AI bots account for roughly 35\% of pageviews but over 65\% of resource-intensive requests, with a preference for low-popularity pages that bypasses regional caches and reaches origin datacenters~\cite{wikimedia2025crawlers}.
Recent proposals have agents retrieve passages that publishers expose under their own policies, in place of whole pages~\cite{chen2025crawled,hua2026semanticsdelivery}.
These works establish the strain and one remedy, but do not explain which agent behaviors drive the load.
We provide that missing layer: passive server-side measurement that attributes origin pressure to specific retrieval properties.

\paragraph{Web proxies.}
Web proxies have long been a major research area.
Caching proxies shared by many users cut latency and load~\cite{rfc9111}, and transcoding or data-saving proxies compress and modify pages to save users' data~\cite{tahir2020weblight,agababov2015flywheel,li2016trafficguard}.
They also raised concerns: measurements found that they removed content and ads and weakened security, and that Web Light applied by default, so publishers could opt out only with a header~\cite{tahir2020weblight,kondracki2020meddling}.
Shared caches have also leaked one user's private page to another, by treating different URLs as the same page~\cite{mirheidari2020cached}; we test this for AI search agents with personalised links and find no leak.
AI search agents combine both roles, but differ in three ways: the user receives an answer and not the page, so neither the user nor the publisher can see what was changed; the processed page is reused across accounts; and the behavior depends on an non-deterministic model, so it can vary between runs.
We measure whether the controls publishers rely on, such as \texttt{Cache-Control}, are respected, observed at the origin server.

\paragraph{Caching in the retrieval stack.}
Most retrieval-augmented systems search a pre-indexed, static corpus, and the caches built for them decide reuse by embedding similarity rather than by URL~\cite{lewis2020rag,bergman2025approximate,jin2026ragcache}.
Work aimed at the open web adds freshness: it reuses a stored answer or page only while it is likely still current, or re-fetches pages and re-indexes only what changed~\cite{mansoor2026semanticcache,zerhoudi2026owlerlite}.
Systems that retrieve from the live web at inference time focus on task completion and answer quality from the agent side~\cite{nakano2021webgpt}.
These works design or propose caches for agents.
We measure the caches that commercial services already run: whether an answer comes from a fetch or from a stored copy, how long that copy is kept, and which \texttt{Cache-Control} signals the service respects, observed at the origin server.
\section{Conclusion}
We measured how AI search agents retrieve the web, using controlled origins and public pages.
Services share fetched pages across accounts and mostly ignore the origin's cache rules, and their fetchers differ in whether they honor retry delays, request \texttt{robots.txt}, or read certain formats, scripts, and long pages.
A publisher therefore cannot predict when a change reaches users, and a user cannot see how old or how complete an answer is.
AI retrieval is a new traffic class: it fetches on demand like a browser, rewrites pages like a transcoding proxy, and reuses them like a shared cache, and the controls HTTP gives publishers do not yet govern it.
Our study covers how these systems behave, not their security; a natural next step is to test the fetchers' security properties, such as TLS validation\cite{kondracki2020meddling}, with the same controlled origins.
\label{BodyEnd}

\bibliographystyle{ACM-Reference-Format}
\bibliography{refs}

\appendix

\section{Ethics}
\label{sec:ethics}

\paragraph{Controlled infrastructure.}
All primary experiments ran on infrastructure we operate: two origins serving static synthetic pages (\origA{} behind Cloudflare's reverse proxy with caching bypassed, \origB{} served directly by Caddy), reached through the vendors' APIs under paid accounts.
The pages contain only synthetic prose and page-specific codes, with no vendor credentials or user data, and no third-party system was stressed or modified: the errors, dropped connections, and redirect loops we injected, and the retries they provoked, landed only on our origins.
Both origins are publicly reachable, so unrelated visitors appear in their logs; we analyze only requests that carry our URL tags.

\paragraph{Terms of service.}
We used OpenAI, Anthropic, and xAI through their published APIs under standard commercial terms, directly or through OpenRouter, with benign web-retrieval prompts within documented rate limits.
For the consumer apps, a researcher entered every question by hand in free, logged-in accounts.
The sharing test used two accounts, one belonging to an author and one to a volunteer; we set up and operated both ourselves and entered only synthetic study prompts.

\paragraph{Public-site validation.}
To check that the controlled findings hold on live pages, we read third-party sites in a way that resembles browsing.
We fetched only logged-out front pages, with no login and no cookies, and kept only visible text; even on sites categorized as social, we fetched the homepage only, never a profile or any personal content, and we collected no personal data.
We kept only sites whose \texttt{robots.txt} allows both our downloader and all four AI page fetchers, and excluded sites that Cloudflare labels as adult, gambling, or CIPA-filtered (\autoref{tab:public-funnel}).
All requests to third-party sites were read-only.
For the file-type test we requested a few announced files per site (\texttt{robots.txt}, sitemaps, feeds, \texttt{llms.txt}, \texttt{.csv} links) from one US server and asked each service about each once, so a site received at most one extra request per service.
For the staleness test we fetched each homepage once every 10 minutes for 24 hours, a rate comparable to an uptime monitor.
For the public-endpoint test, we read each of eight public counters once every five minutes.
The downloader identified itself as an academic research tool and kept only each counter's value, not the response.
We release only these values and their times, and each vendor's answer, with Anthropic's reported fetch time.

\paragraph{Vendor attribution.}
We attribute requests to vendors by a tag in the URL we give each vendor, matched only against our own server logs.
We also record the fetchers' source addresses, only to count them, look up the network that owns each, and compare them with vendor-published lists, never to identify an individual.

\paragraph{Responsible disclosure.}
We measure vendors' systems from the outside, through their public APIs and apps.
Our findings describe how their fetchers treat publishers and users, and are not security vulnerabilities, so we did not use a coordinated-disclosure window.

\paragraph{IRB}
This study collects no personal data from anyone: its data come from our own origins, the vendors' APIs, and logged-out public pages.
The one volunteer account in the sharing test was set up and operated by us, and we kept no data about its owner.
We asked our institution's review board, which confirmed that the study involves no human data or samples and does not require review.

\section{Use of Generative AI}
\label{app:genai}
We used generative AI for two tasks: rewriting text, and writing code.
We used Claude Code (Anthropic) with Claude Opus 4.8, Opus 5, Opus 5.5, and Sonnet 5.5 between September and November 2026.

For text, we wrote the outline and the claims of each section and the tool reworded passages.
We revised the text and checked every claim against our data.
For code, it wrote most of the measurement harness, the test pages, the analysis scripts, and the figure code, set up the two origins, and prepared the released code and data.

The study was exploratory: we chose each next measurement from the previous findings, and we checked every output.
For example, we first saw that pages are shared across accounts and that failed fetches behave oddly; we then measured how long a copy lasts, and then whether this differs by file format.
Each new measurement also served as a cross-check of the earlier ones.

\section{Test Materials}
\label{app:test-materials}
This appendix gives the materials behind the measurement setup (\autoref{sec:setup}): the prompts we sent (\autoref{app:prompts}, \autoref{app:prompts-public}), the public pages we chose (\autoref{app:public-sites}, \autoref{app:public-pilot}), how we built the test documents (\autoref{app:test-documents}), and how we ran the tests (\autoref{app:openrouter}, \autoref{app:reuse-design}).

\subsection{Prompts for Controlled Origins}
\label{app:prompts}

We use one prompt shape for nearly all tests on our own origins, so that the same question can be reused across services and experiments.
The question asks for a code that the page contains, so every answer can be checked deterministically by a script.

\para{Controlled-origin API prompts.}
Every API trial sent only the tested URL and one question:

\begin{verbatim}
<URL>

Fetch this document and tell me its id and the code
beginning with TRUNC- for the section at depth=N%
in it. Quote them exactly.
\end{verbatim}

Only the depth $N$ changed between trials.
Each depth code appears on one page only, so a correct answer shows that the model read that page.

\parai{Exceptions.}
Three experiments ask a different question, because they need a different answer:
\begin{itemize}
  \item \emph{Document size:} every code visible in the page, to see how far down the model read.
  \item \emph{Faults:} what the page said and any visible codes, so the model reports the failure it received.
  \item \emph{URL equivalence:} one entry from a register of numbered entries on the page, which shows which page the answer came from.\footnote{The test checks whether different URLs count as the same page, so the URL cannot carry the tag that separates pages in other experiments.}
\end{itemize}

\parai{Prompt wording.}
We also ran with differently worded prompts, and no fetcher's behavior changed.

\subsection{Prompts for Public Pages}
\label{app:prompts-public}
Public pages carry no depth codes, so the standard question could not apply.
Instead, we asked for the text at a fixed place on the page, which we could check word by word against our own downloads.
The prompt named a line on the page, the \emph{anchor}, and asked for the words that follow it:

\begin{verbatim}
Fetch this document: <URL>. Quote exactly the first
12 words that follow the line "<ANCHOR>" in it.
If you cannot read the document, say so plainly.
\end{verbatim}

The anchor was a line that stayed the same across our downloads of the page, while the text after it changed, so our downloads told us which version of the page each answer came from.
We ask for 12 words because longer quotations can be refused: Google documents a \texttt{RECITATION} stop reason for content that may be copied from a source~\cite{google2026troubleshooting}, and in our runs OpenAI's model declined to quote more than 25 words verbatim.\footnote{This limit comes from the model's own reply, for example ``I can't reproduce more than 25 words verbatim from a non-lyrical source''.
We found no OpenAI document that states it.
The OpenAI Model Spec only states a general duty to respect creators' rights~\cite{openai2026modelspec}.}
For pages written without spaces between words, such as Japanese or Chinese, the prompt asked for the first 30 characters instead.

\para{Public endpoints.}
The staleness test on the eight endpoints asked for the current value of a field that only increases:

\begin{verbatim}
Fetch <URL> and reply with ONLY <TARGET>, exactly
as shown, and nothing else, no words, no
punctuation, just the value.
\end{verbatim}

\para{File types.}
The file-type test asked whether a service could read a public file of a given type, by asking it to quote the file's first three lines:

\begin{verbatim}
Fetch this document: <URL>. Quote its first 3 lines
exactly. If you cannot read the document, say so
plainly.
\end{verbatim}

\subsection{Public Sites}
\label{app:public-sites}

A site is eligible if we may ask the services about it; each test then selects what it needs from the eligible sites.
The public checks use three sets of targets (\autoref{sec:setup-targets}): public files and changing pages, selected from the eligible sites, and counter endpoints, chosen by hand.

\para{Site sample.}
We draw sites from the Chrome User Experience Report (CrUX)~\cite{google2017crux}, using the August 2026 release of its global top list as published by the CrUX top-lists repository~\cite{durumeric2022cruxtoplists, ruth2022toplists}.
We take all 1,000 sites of the top group, plus 1,000 sites sampled at random from the next group (ranks 1,001--5,000), for 2,000 sites in total.

\begin{table}[t]
  \centering
  \small
  \begin{tabular}{lr}
    \toprule
    Step & Sites \\
    \midrule
    CrUX sample & 2,000 \\
    \texttt{robots.txt} lets our downloader in & 1,783 \\
    Normal answer (HTTP 200) from the US server & 1,272 \\
    \texttt{robots.txt} allows all four AI page fetchers & 1,179 \\
    Not classed as adult by Cloudflare's family DNS & 858 \\
    No excluded URL Scanner category & \textbf{776} \\
    \bottomrule
  \end{tabular}
  \caption{Sites remaining after each step.}
  \label{tab:public-funnel}
\end{table}

\para{Eligibility.}
A site is eligible if it answers normally, its \texttt{robots.\allowbreak txt} lets our downloader and the page fetchers of all four services in, and its content is not sensitive (\autoref{tab:public-funnel}).
The \texttt{robots.\allowbreak txt} condition means that we only ask the services about pages that the site allows to be fetched; we do not rely on the fetchers to stay away, because some user-triggered ones may not honor it (OpenAI says so for \texttt{ChatGPT-User}~\cite{openai2026bots}).
We remove sensitive content in two stages.
First, Cloudflare's family DNS resolver (1.1.1.3), which answers \texttt{0.0.0.0} for domains Cloudflare classes as adult content, removes those sites without visiting them.
Second, following Habib et al.~\cite{habib2026fetch}, we take each remaining site's categories from Cloudflare's URL Scanner and exclude any site labelled \emph{CIPA}, \emph{CIPA Filter}, \emph{Adult Themes}, \emph{Pornography} or \emph{Gambling}.
When a homepage redirects, we use the labels of the page it lands on.
Sites Cloudflare leaves unlabelled are kept.

\para{Reference copies.}
In every public test we compare each answer with our own downloads of the target file or page, taken around the time we asked (for the counter endpoints, every five minutes throughout).

\subsubsection{Public Files}
\label{app:public-files}

The file-type test asks whether each service can read a public file of a given declared type, by asking it to quote the file's first three lines (\autoref{app:prompts-public}).

\paragraph{Selection}
On the eligible sites we look only where sites announce files: sitemaps listed in \texttt{robots.txt} (or at \texttt{/sitemap.\allowbreak xml}), RSS and Atom feeds linked from the homepage, \texttt{/llms.\allowbreak txt}, and links to \texttt{.csv} files.
We classify each file by the type the server declares in its \texttt{Content-Type} header, not by its name or content, because the fetchers accept or reject a file based on that declared type (for example, OpenAI's fetcher reports ``Unsupported content-type: application/xml'', \autoref{sec:content-type-filtering}).
We keep files that answered normally, are 1--200\,KB with at least five lines, whose host's \texttt{robots.txt} allows all four fetchers, and that the server provides in the file format we request (\autoref{tab:file-funnel}).
Most files dropped by the type step were served as \texttt{text/html}.
We also removed one site by hand, because its domain name indicates adult content although neither filter had flagged it.
This leaves 299 files: 226 XML and, for comparison, 73 \texttt{llms.txt} files served as \texttt{text/plain}.

\begin{table}[t]
\centering
\footnotesize
\caption{Public files: files remaining after each step.
Sitemaps and feeds must declare an XML type, and \texttt{llms.txt} files must declare \texttt{text/plain}.}
\label{tab:file-funnel}
\setlength{\tabcolsep}{3pt}
\begin{tabular}{@{}L{1.65in}rrrr@{}}
\toprule
Step & Sitemap & Feed & \texttt{llms.txt} & Total \\
\midrule
Candidate checked & 776 & 141 & 776 & 1,693 \\
Answered normally (HTTP 200) & 511 & 102 & 213 & 826 \\
1--200\,KB, at least 5 text lines & 240 & 57 & 165 & 462 \\
\texttt{robots.txt} allows all four fetchers & 239 & 57 & 165 & 461 \\
Declared a requested type & 171 & 55 & 73 & \textbf{299} \\
\bottomrule
\end{tabular}
\end{table}

\paragraph{Markdown and CSV.}
These two types are almost absent from the public web, so we check them on our controlled origins only.

In detail, only one file was declared \texttt{text/markdown} (but over our size limit), none was CSV, and no site served a Markdown homepage at \texttt{/index.md} or \texttt{/index.html.md}, the addresses the \texttt{llms.txt} proposal defines~\cite{howard2024llmstxt}.
Twelve homepages (1.5\%) return Markdown only when a request asks for it (\texttt{Accept: text/markdown}), as offered by Cloudflare's ``Markdown for Agents''~\cite{cloudflare2026markdownagents}.
However, none of the fetchers we could identify asks for content specifically in Markdown.\footnote{On our own origins, the fetchers of OpenAI, Anthropic and Google asked for any format (\texttt{*/*}) or for HTML first, never for Markdown; xAI's fetcher presents itself as a browser and could not be singled out.
We cannot observe requests on the public sites themselves.}

\paragraph{Test run.}
We asked each service once per file, spacing calls three minutes apart per service.

\subsubsection{Changing Pages}
\label{app:public-pilot}
The staleness test needs two things from each page: a part that changes often, and a way to point at it.
We can only ask for a short quotation (\autoref{app:prompts-public}), so we cannot always start from the top of the page; the prompt instead names an \emph{anchor}, a short line just above the changing part.
The test itself is in \autoref{app:public-staleness}.

\paragraph{Watching.}
We visited the homepage of each of the 776 eligible sites every 10 minutes for 24 hours.
Per visit we stored the status, the response headers without cookies, the sizes and the visible text, but not the HTML.
An \emph{anchor} is a short line just above the changing part of the page that appears exactly once in at least 90\% of our visits.
A \emph{real change} is a new line that is not a one-off: it is still on the page at the next visit, so it lasts at least 10 minutes.
We ignore clock and date lines, and lines with too few letters or with at least 30\% digits; letters count in any script.
In a sample of 30 changes, 70\% were real.

\paragraph{Selection.}
We keep a site if it serves HTML, answered at least 100 visits, changes often enough (a median gap between real changes under 30 minutes),\footnote{This is below the reuse time we observed for OpenAI (\autoref{sec:retention-and-renewal}), so a reused copy can be told from a current one.} and showed a real change in at least 30\% of the time.
This leaves 172 sites, of which 90 have a stable anchor thus become our test targets (\autoref{tab:changing-funnel}).

\begin{table}[t]
\centering
\small
\caption{Changing pages: sites remaining after each step.}
\label{tab:changing-funnel}
\begin{tabular}{lr}
\toprule
Step & Sites \\
\midrule
Sites watched for 24 hours & 776 \\
At least one normal answer (HTTP 200 with text) & 772 \\
Serves HTML & 769 \\
At least 100 normal answers & 763 \\
Median gap between real changes under 30 minutes & 196 \\
Real change in at least 30\% of the 10-minute steps & 173 \\
Personal-data exclusion & 172 \\
Stable anchor & \textbf{90} \\
\bottomrule
\end{tabular}
\end{table}

\subsubsection{Counter Endpoints}
\label{app:counter-endpoints}

\paragraph{Selection.}
These are eight public endpoints whose values only increase, such as the largest Hacker News item ID, a blockchain height, or the USGS feed's \texttt{metadata.generated} timestamp.
Few public services expose such a value, so we chose them by hand, keeping those that need no login and return the value as plain text or JSON.
The set covers several cache headers and hosts, so a header's effect can be told apart from its host's (\autoref{tab:anchor-staleness}).

\paragraph{Test run.}
We asked each service for each value every 45 minutes for about three days.
Staleness is how out of date an answer is.
That is, if the value a service gave matches our reference log recorded two hours ago, the answer is two hours stale.
For Anthropic, whose reply states when the page was retrieved, we use the age of that copy.\footnote{We checked the reported time against our reference log, and the two agree to within about a minute.}
However, our reference log stores only the numeric value, so staleness below five minutes cannot be resolved.

\subsection{Test Document Construction}
\label{app:test-documents}

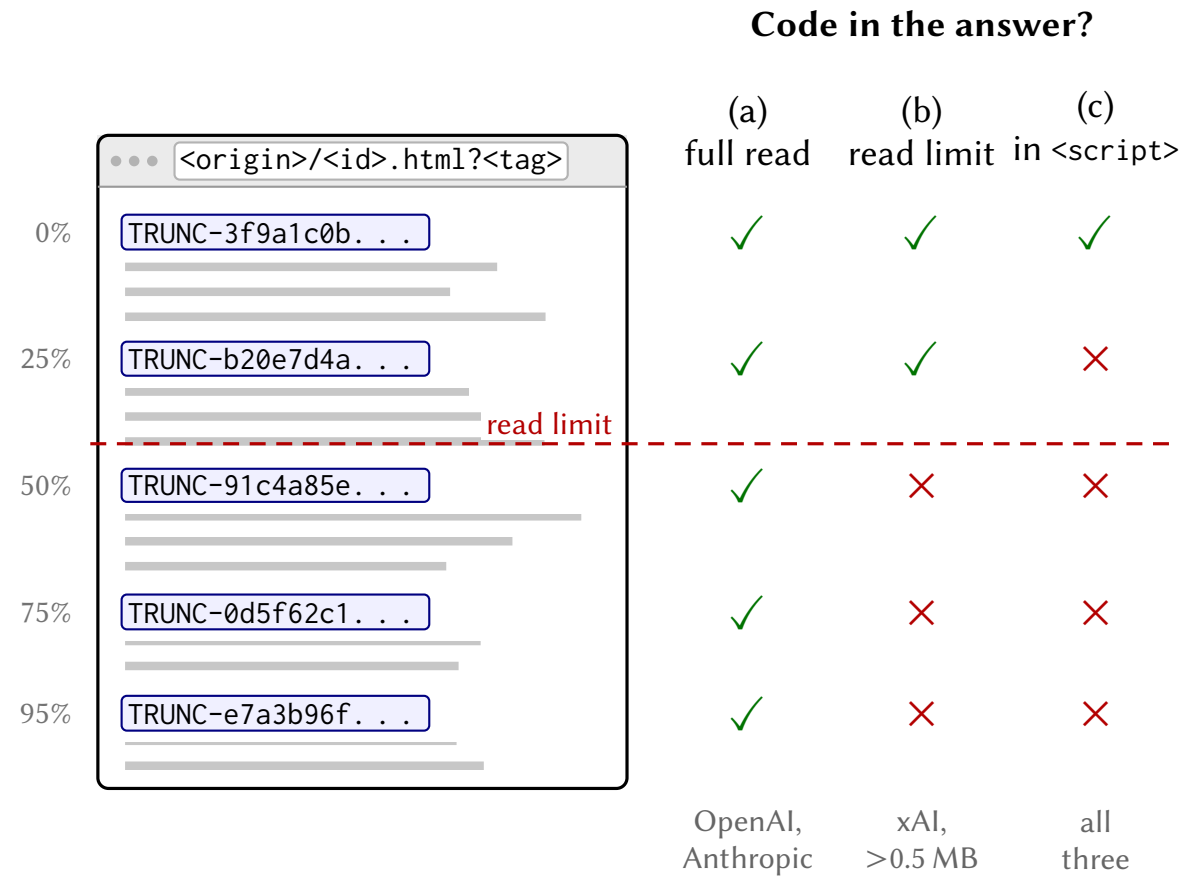
\begin{figure}[t]
\centering
\begin{tikzpicture}[font=\scriptsize\sffamily,
  code/.style={draw=blue!50!black, fill=blue!6, rounded corners=1pt,
               inner sep=1.2pt, font=\tiny\ttfamily, anchor=west},
  ok/.style={text=green!45!black, font=\small\bfseries},
  miss/.style={text=red!70!black, font=\small\bfseries},
  hdr/.style={align=center, font=\scriptsize\sffamily, anchor=south},
  who/.style={align=center, font=\tiny\sffamily, text=black!60, anchor=north}]
\def\W{3.5}\def\H{3.9}
\fill[black!8] (0,0.42) rectangle (\W,0.08);
\foreach \i in {0,1,2} \fill[black!30] (0.12+0.12*\i,0.25) circle (0.035);
\node[draw=black!30, fill=white, rounded corners=1pt, inner sep=1pt,
      font=\tiny\ttfamily, anchor=west] at (0.5,0.25)
  {<origin>/<id>.html?<tag>};
\foreach \i in {0,...,22} {
  \pgfmathsetmacro{\y}{-0.12-0.165*\i}
  \pgfmathsetmacro{\L}{2.75+0.45*sin(\i*97)}
  \draw[black!22, line width=1.6pt] (0.18,\y) -- (\L,\y);
}
\foreach \d/\c/\n in {0/3f9a1c0b/0, 25/b20e7d4a/1, 50/91c4a85e/2, 75/0d5f62c1/3, 95/e7a3b96f/4} {
  \pgfmathsetmacro{\y}{-0.22-\d/100*(\H-0.55)}
  \fill[white] (0.05,\y-0.19) rectangle (\W-0.05,\y+0.19);
  \node[code] at (0.15,\y) {TRUNC-\c\ldots};
  \node[text=black!55, anchor=east, font=\tiny\sffamily] at (-0.05,\y) {\d\%};
  \coordinate (r\n) at (0,\y);
}
\draw[line width=0.6pt, rounded corners=2pt] (0,0.42) rectangle (\W,-\H);
\draw[line width=0.4pt, black!40] (0,0.08) -- (\W,0.08);
\draw[red!70!black, densely dashed, line width=0.6pt] (-0.05,-1.62) -- (7.1,-1.62);
\node[red!70!black, font=\tiny\sffamily, anchor=south east, fill=white, inner sep=1pt]
  at (\W-0.05,-1.6) {read limit};
\def\xa{4.3}\def\xb{5.45}\def\xc{6.6}
\node[hdr] at (\xa,0.1) {(a)\\full read};
\node[hdr] at (\xb,0.1) {(b)\\read limit};
\node[hdr] at (\xc,0.1) {(c)\\in \texttt{\tiny<script>}};
\node[font=\scriptsize\sffamily\bfseries, anchor=south] at (\xb,0.95) {Code in the answer?};
\foreach \n in {0,...,4} \node[ok] at (\xa,0|-r\n) {\cmark};
\foreach \n in {0,1} \node[ok] at (\xb,0|-r\n) {\cmark};
\foreach \n in {2,3,4} \node[miss] at (\xb,0|-r\n) {\xmark};
\node[ok] at (\xc,0|-r0) {\cmark};
\foreach \n in {1,...,4} \node[miss] at (\xc,0|-r\n) {\xmark};
\node[who] at (\xa,-\H-0.02) {OpenAI,\\Anthropic};
\node[who] at (\xb,-\H-0.02) {xAI,\\${>}$0.5\,MB};
\node[who] at (\xc,-\H-0.02) {all\\three};
\end{tikzpicture}
\caption{Codes at known positions show which parts of a page reach the answer.
A longer test page carries five codes at 0--95\% of its text, each unique to the page.
(a)~OpenAI and Anthropic return every code up to 2\,MB.
(b)~xAI returns only the codes within the first 0.5\,MB of a larger page (\autoref{sec:token-scaling}).
(c)~When the codes after the first sit inside \texttt{<script>} tags, no service returns them; the 0\% code, plain HTML, shows the page itself was read (\autoref{sec:javascript}).}
\label{fig:known-codes}
\end{figure}

\para{Source content.}
Our origin generates each test document on request from ordered source lines: short pages for tests that need one value from the page, and longer pages for tests that need to find information at different depths.
Unless an experiment says otherwise, it serves a 2\,KB HTML page.
A longer page holds prose from a frozen corpus, with five codes placed near 0\%, 25\%, 50\%, 75\%, and 95\% of the text.
The codes show whether an answer comes from the intended page and, for a longer page, from the requested position.
Each code is an HMAC of a secret shared by the origin and the measurement client, the page ID, the code's location on the page, and a content revision.
Independent trials use fresh page IDs.
When a test changes a page in place, the revision changes too, so answers from the old and the new content can be distinguished.
\autoref{fig:known-codes} shows how the codes reveal which parts of a page reached the answer.

\para{Formats.}
The origin renders the same source lines in eight document formats (\autoref{lst:test-document-renderers}).
Format-specific markup changes the bytes and the file size, but not the codes or their order.
We checked with independent tools that each output can be opened and its codes extracted (\autoref{tab:test-document-validation}).

\begin{lstlisting}[
  caption={Rendering the ordered source lines into test documents.
  Each renderer preserves the page-specific codes and their order.},
  label={lst:test-document-renderers},
  basicstyle=\ttfamily\footnotesize,
  columns=fullflexible,
  breaklines=true
]
input: source_lines  // prose and codes in order

text:
    join(source_lines, "\n") + "\n"

markdown:
    "# probe document\n\n"
      + join(source_lines, "\n\n") + "\n"

html:
    write doctype, charset, and title
    for line in source_lines:
        write "<p>" + html_escape(line) + "</p>"

json:
    json_encode({"document": source_lines})

xml:
    write XML header and "<document>"
    for line in source_lines:
        write "<line>" + xml_escape(line)
                       + "</line>"
    write "</document>"

csv:
    write_csv_row(["n", "content"])
    for index, line in enumerate(source_lines):
        write_csv_row([index, line])

pdf:
    write PDF header, catalog, page tree,
          and Helvetica font
    for page_lines in groups_of_55(source_lines):
        stream = begin_text()
        for line in page_lines:
            stream += show_text(pdf_escape(line))
            stream += next_line()
        stream += end_text()
        write page object and its text stream
    write cross-reference offsets and trailer

docx:
    document_xml = new_word_document()
    for line in source_lines:
        document_xml += word_paragraph(
            xml_escape(line))
    zip_with_fixed_timestamps(
        "[Content_Types].xml",
        "_rels/.rels",
        "word/document.xml" = document_xml)
\end{lstlisting}

\para{JavaScript test pages.}
These pages test whether a service reads content that a script delivers.
They come in five versions (\autoref{tab:js-variants}), each built from a fresh page ID and returning a 4\,KB page with five codes and the standard question (\autoref{app:prompts}).
The 0\% code is always plain HTML as a control, so its return shows that the page reached the model.
The other four codes (25--95\%) are delivered differently in each version (\autoref{fig:js-source}, \autoref{alg:js-pages}); the scrambled ones can be read only by running the page's decoder.
Headless Chromium shows all five codes on every version except JSON-LD, which browsers never display.
We ran three sittings, each with a fresh page for every service and version, using OpenAI, xAI, and Anthropic's basic fetch tool (\texttt{web\_fetch\_20250910}).

\begin{figure}[t]
\lstset{language=HTML, basicstyle=\ttfamily\footnotesize, columns=fullflexible,
        breaklines=true, frame=single, aboveskip=2pt, belowskip=2pt}
\begin{lstlisting}[title={\textbf{Plain HTML}}]
<p>TRUNC-3f9a1c0b7d2e</p>
\end{lstlisting}
\begin{lstlisting}[title={\textbf{Data block}}]
<script type="application/json">["TRUNC-3f9a1c0b7d2e"]</script>
<script> ...adds each line to the page... </script>
\end{lstlisting}
\begin{lstlisting}[title={\textbf{JSON-LD}}]
<script type="application/ld+json">
  {"@type": "Article", "articleBody": "TRUNC-3f9a1c0b7d2e"}
</script>
\end{lstlisting}
\begin{lstlisting}[title={\textbf{Inline script (scrambled)}}]
<script> seed = 1234567890; blob = "Xk9a..."; ...decode and add to the page... </script>
\end{lstlisting}
\begin{lstlisting}[title={\textbf{External script (scrambled)}}]
<script src="/js/p/n.js"></script>
\end{lstlisting}
\caption{Page source of each version of a JavaScript test page, shortened to one code.
A browser shows the text on every version except JSON-LD.
The 0\% code, a plain paragraph in every version, is left out.}
\label{fig:js-source}
\end{figure}

\begin{algorithm}[t]
\small
\DontPrintSemicolon
\SetKwProg{Fn}{function}{:}{}
\SetKwFunction{FScramble}{scramble}
\SetKwFunction{FDecoder}{decoder}
\caption{Building a JavaScript test page.
All versions share the setup and differ only in how the page body is built from \textit{test\_lines}.}
\label{alg:js-pages}
\KwIn{page id $p$, size $n$, version $v$}
$\textit{test\_lines} \gets$ $n$ bytes of each code at 25, 50, 75, 95\%\;
\BlankLine
\tcc{Unscrambled, readable in plain text}
\uIf{$v$ = plain}{
  $\textit{body} \gets$ \texttt{<p>}\textit{line}\texttt{</p>} for each \textit{line} in \textit{test\_lines}\;
}
\uElseIf{$v$ = data block}{
  $\textit{body} \gets$ \texttt{<script type="application/json">} json(\textit{test\_lines}) \texttt{</script>} + a script that copies it into the page\;
}
\uElseIf{$v$ = JSON-LD}{
  $\textit{body} \gets$ \texttt{<script type="application/ld+json">} json(Article, articleBody $=$ \textit{test\_lines}) \texttt{</script>}\;
}
\Else{
  \tcc{Scrambled, readable by running decoder}
  $\textit{text} \gets$ join(\textit{test\_lines}, newline)\;
  $\textit{seed} \gets$ first 4 bytes of SHA-256($p$, $n$)\;
  $\textit{blob} \gets$ base64(\FScramble{\textit{text}, \textit{seed}})\;
  $\textit{js} \gets$ \FDecoder{\textit{seed}, \textit{blob}}\;
  \eIf{$v$ = inline}{
    $\textit{body} \gets$ \texttt{<script>} \textit{js} \texttt{</script>}\;
  }{
    $\textit{body} \gets$ \texttt{<script src="/js/p/n.js"></script>}\;
    serve \textit{js} at \texttt{/js/p/n.js}\;
  }
}
\BlankLine
\Fn{\FScramble{\textit{text}, \textit{seed}}}{
  $s \gets \textit{seed}$\;
  \ForEach{byte $b$ in \textit{text}}{
    $s \gets$ xorshift32($s$)\;
    output $b \oplus (s \bmod 256)$\;
  }
}
\Fn{\FDecoder{\textit{seed}, \textit{blob}}}{
  \tcc{runs in the browser}
  $\textit{text} \gets$ \FScramble{base64\_decode(\textit{blob}), \textit{seed}}\;
  \ForEach{line in split(\textit{text}, newline)}{
    append \textsf{paragraph}(line) to the page\;
  }
}
\end{algorithm}

\begin{table}[t]
\centering
\small
\caption{Tools that check each generated format opens and yields its codes.
We checked a short page and a 64\,KiB five-code page in each format.}
\label{tab:test-document-validation}
\begin{tabular}{@{}lp{4.8cm}@{}}
\toprule
Format & Validation tools \\
\midrule
HTML & Python \texttt{html.parser} \\
Text, Markdown & Strict UTF-8 decoding \\
JSON & Python \texttt{json} \\
XML & Python \texttt{xml.etree} \\
CSV & Python \texttt{csv}; row-width check \\
PDF & Poppler, MuPDF, pypdf, pdfminer.six \\
DOCX & python-docx, docx2txt, LibreOffice;
       ZIP and XML checks \\
\bottomrule
\end{tabular}
\end{table}

\subsection{OpenRouter and Direct Routes}
\label{app:openrouter}
OpenRouter is a gateway that reaches many vendors' models through one API.
In principle, it would let us extend the API tests to more vendors, but it can replace a vendor's fetcher with its own, which would measure OpenRouter and not the vendor.
We therefore check that the fetch comes from the vendor's own tool, by the User-Agent of the request, since OpenRouter's fetcher uses its own User-Agent.\footnote{We use the same URL tag as elsewhere (\autoref{sec:setup-probes}): it exists only in one question, so every request for that URL at our origin ties to that question.}

For OpenAI, xAI, and Anthropic's \texttt{claude-sonnet-5}, the fetch came from the vendor's own tool, and OpenRouter's fetcher never appeared.
The check is per model: \texttt{claude-haiku-4.5} and DeepSeek both got OpenRouter's fetcher, while Sonnet got Anthropic's own, so we use Sonnet and do not query DeepSeek.
OpenRouter's documentation lists the models for which it uses the vendor's own web tool, and neither is on that list~\cite{openrouter2026websearch}.
This is why the controlled-origin API tests cover only OpenAI, Anthropic, and xAI.

\subsection{Reuse Test Design}
\label{app:reuse-design}

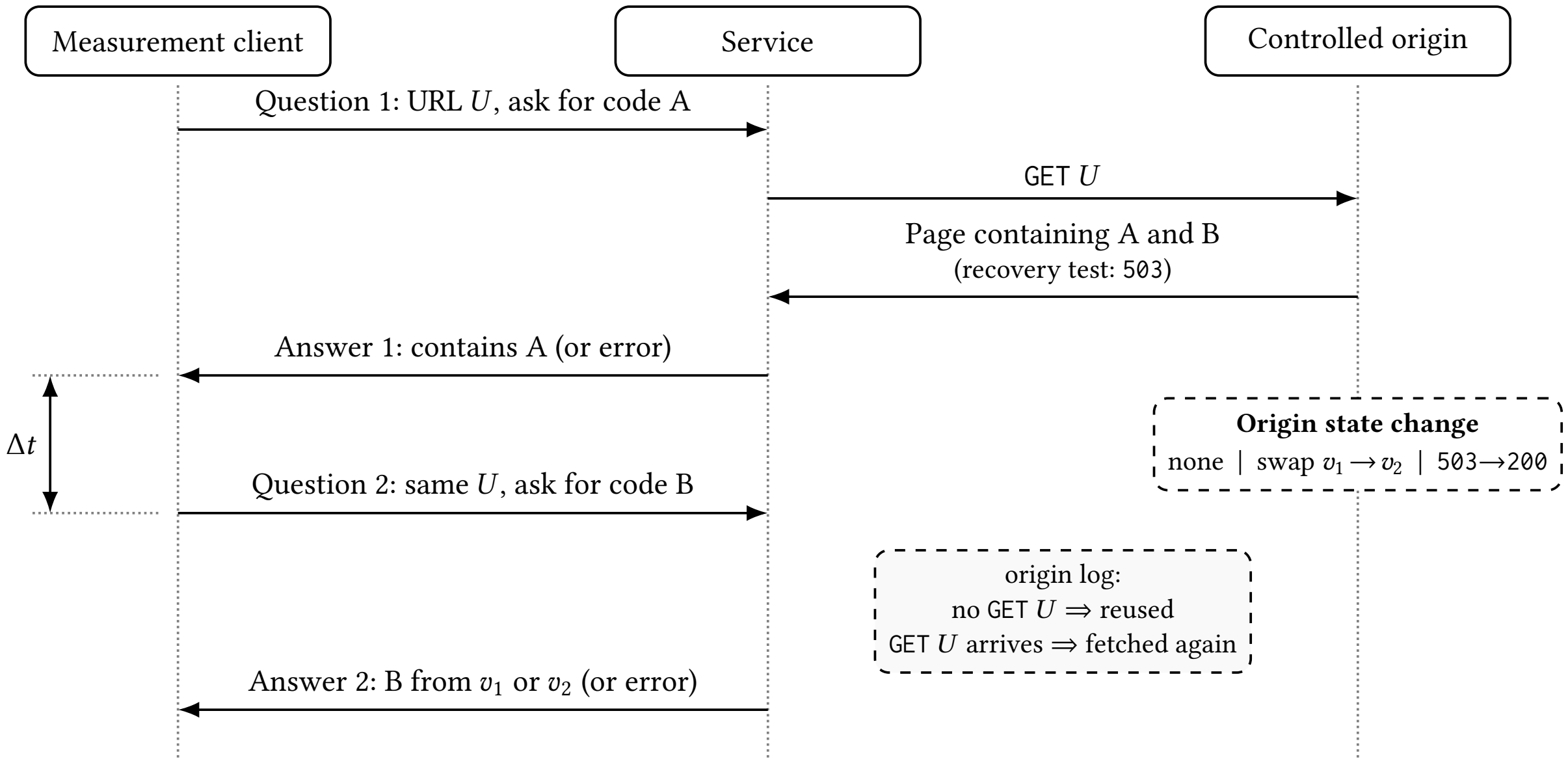
\begin{figure*}[t]
\centering
\begin{tikzpicture}[
  >=Latex,
  font=\small,
  line width=0.7pt,
  participant/.style={
    draw, rounded corners, align=center,
    minimum width=3.1cm, minimum height=0.7cm
  },
  note/.style={
    draw, dashed, rounded corners, fill=gray!5,
    font=\footnotesize, align=center, inner sep=4pt
  }
]
\node[participant] at (0,0)  {Measurement client};
\node[participant] at (6,0)  {Service};
\node[participant] at (12,0) {Controlled origin};

\draw[densely dotted, gray] (0,-0.45)  -- (0,-7.3);
\draw[densely dotted, gray] (6,-0.45)  -- (6,-7.3);
\draw[densely dotted, gray] (12,-0.45) -- (12,-7.3);

\draw[->] (0,-0.9) -- node[above] {Question 1: URL $U$, ask for code A} (6,-0.9);
\draw[->] (6,-1.6) -- node[above] {\texttt{GET} $U$} (12,-1.6);
\draw[->] (12,-2.6) -- node[above, align=center]
  {Page containing A and B\\[-1pt]{\footnotesize (recovery test: \texttt{503})}} (6,-2.6);
\draw[->] (6,-3.4) -- node[above] {Answer 1: contains A (or error)} (0,-3.4);

\node[note, fill=white] (state) at (12,-4.1)
  {\textbf{Origin state change}\\[1pt]
   none \,$\mid$\, swap $v_1\!\to\!v_2$ \,$\mid$\, \texttt{503}$\to$\texttt{200}};

\draw[densely dotted, gray] (-0.2,-3.4) -- (-1.5,-3.4);
\draw[densely dotted, gray] (-0.2,-4.8) -- (-1.5,-4.8);
\draw[<->] (-1.3,-3.4) -- node[left] {$\Delta t$} (-1.3,-4.8);

\draw[->] (0,-4.8) -- node[above] {Question 2: same $U$, ask for code B} (6,-4.8);

\node[note] at (9,-5.8)
  {origin log:\\
   no \texttt{GET} $U$ $\Rightarrow$ reused\\
   \texttt{GET} $U$ arrives $\Rightarrow$ fetched again};

\draw[->] (6,-6.8) -- node[above] {Answer 2: B from $v_1$ or $v_2$ (or error)} (0,-6.8);
\end{tikzpicture}

\caption{Reuse test design.
A first question makes the service fetch a tagged URL $U$.
After a gap $\Delta t$, the origin keeps the page (reuse, \autoref{sec:content-reuse-and-sharing}), swaps it for a new version (update, \autoref{sec:retention-and-renewal}), or turns a \texttt{503} into a marked \texttt{200} page (recovery, \autoref{sec:neg-cache}).
A second question asks for a code absent from the first answer; the origin log shows whether the service fetched again.}
\label{fig:reuse-setup}
\end{figure*}

Every caching test in \autoref{sec:cache} follows one pattern (\autoref{fig:reuse-setup}):
A first question makes the service fetch a tagged page; after a delay, during which the origin changes the page or does nothing, a second question asks for a code that only a refetch could return.
The origin log and the answer together show whether the service reused its earlier retrieval.

\section{Additional Fetcher and Content Results}
\label{app:fetcher-content}
This appendix supports \autoref{sec:behavior} and \autoref{sec:content}: where the fetchers connect from (\autoref{app:source-networks}), how they respond to faults (\autoref{app:fault-behavior}, \autoref{app:fault-tokens}), handle \texttt{robots.txt} (\autoref{app:robots-apps}, \autoref{app:robots-reuse}), and read files and pages (\autoref{app:public-xml}, \autoref{app:gemini-origins}).

\subsection{Where the Fetchers Connect From}
\label{app:source-networks}

This appendix shows which networks the fetchers' requests come from (\autoref{sec:fetcher-id}).
We tie each request to a vendor by the tag in its URL.
For OpenAI and Anthropic, the request must also carry that vendor's crawler name; xAI's fetcher sends a browser User-Agent, so for xAI the tag alone decides.
For each request, we looked up the network that owns the source address in a public ownership database~\cite{iptoasn2026database}, offline.

OpenAI's requests come from one network, owned by Microsoft, and Anthropic's from two, owned by Google and Amazon.
xAI uses a new address for almost every request (7{,}774 addresses for 8{,}286 requests), spread over about 1{,}260 networks, which is not how a fixed set of servers behaves.

Those networks are too many to name, so we use ASdb~\cite{ziv2021asdb}, a public database that gives each network a business category.
It files 82\% of xAI's requests under ``Internet Service Provider'' and 7\% under hosting.
ASdb labels a network by its owner's function\footnote{It files Microsoft's network under software development and Google's under search, although both are cloud networks.}, so we also labeled a random sample of 100 xAI requests by hand with ASdb's categories, using the owner's website and public pages about the network for unclear cases.
38 came from hosting providers and 62 from internet or phone providers, including 12 from one backbone carrier.
About half of the providers are in South America.
Many telecom companies also sell hosting alongside access, so we cannot separate the two exactly.

\subsection{Origin Responses Under Faults}
\label{app:fault-behavior}

We served each fault from our origins in response to one question and counted the requests that arrived for it (\autoref{tab:fault-behavior-full}).
A number in parentheses is the count of requests per question, ranges span six runs (three on each origin), and ``Overlap'' marks requests that were concurrent, so they cannot all be retries.
OpenAI repeats requests after server errors and dropped connections, Anthropic sends one in nearly every case, and xAI sends overlapping requests whatever the fault.
Two faults differ by origin:
\begin{smartlist}
\item{Cut-off response.} Our server ends the response before the page is complete. On \origA{}, Cloudflare answers the fetcher with its own HTTP 502 and the fetchers sent more requests after it; on \origB{}, the fetchers received the incomplete page and discarded it.
\item{Dropped connection.} The connection is closed with no response. The fetcher gets an HTTP 502 on both origins, from Cloudflare on \origA{} and from Caddy on \origB{}.
\end{smartlist}
This is the full version of \autoref{tab:fault-behavior} in \autoref{sec:failure-retries}, which shows the subset behind the main findings.

\begin{table}[t]
\centering
\small
\caption{Requests that reached our origin per question, for every fault we injected (full version of \autoref{tab:fault-behavior}).
$^*$: the requests were concurrent, so they cannot all be retries.
In the redirect-loop row, each arrival means one hop.}
\label{tab:fault-behavior-full}
\begin{tabular}{@{}lrrr@{}}
\toprule
Origin response & OpenAI & Anthropic & xAI \\
\midrule
\multicolumn{4}{@{}l}{\emph{Responses with \texttt{Retry-After}}} \\
503, delay 10\,s        & 3--4 & 1    & 5--7$^*$ \\
429, delay in seconds   & 1    & 1    & 5--7$^*$ \\
429, delay as date      & 1    & 1    & 6--7$^*$ \\
\midrule
\multicolumn{4}{@{}l}{\emph{Errors and connections}} \\
403                     & 1    & 1    & 5--7$^*$ \\
404                     & 1    & 1    & 4--7$^*$ \\
500                     & 3--4 & 1    & 5--7$^*$ \\
Connection dropped      & 3--5 & 1--3 & 6--8$^*$ \\
Response cut off        & 1--2 & 1--4 & 6--28$^*$ \\
\midrule
\multicolumn{4}{@{}l}{\emph{Unusual successes and recovery}} \\
200 after 25\,s         & 1    & 1--3 & 7$^*$ \\
Empty 200               & 1    & 1    & 5--7$^*$ \\
Two failures, then 200  & 3    & 1    & 5--7$^*$ \\
\midrule
\multicolumn{4}{@{}l}{\emph{Redirects}} \\
Redirect loop           & 2    & 10   & 57--77$^*$ \\
\bottomrule
\end{tabular}
\end{table}

\subsection{Consumer-App \texttt{robots.txt} Requests}
\label{app:robots-apps}

We applied the origin-level check from \autoref{sec:robot-txt-compliance} to the consumer chat apps: did the app request \texttt{robots.txt} before its first tagged-page request on its test origin (\autoref{tab:robots-apps})?
A \texttt{robots.txt} request cannot carry our URL tag, so we cannot tie it to a question; we attribute it to an app by its User-Agent and address, and ``No'' means that no request was attributed to the app.

\begin{table}[t]
\centering
\small
\caption{Did the app request \texttt{robots.txt} before its first tagged-page request on its test origin?}
\label{tab:robots-apps}
\begin{tabular}{@{}ll@{}}
\toprule
App & Read first? \\
\midrule
Mistral    & Yes \\
Gemini     & No \\
Grok       & No \\
DeepSeek   & No \\
Qwen       & No \\
ChatGPT    & No$^\dagger$ \\
Claude     & Unclear$^\dagger$ \\
Meta AI    & Unclear$^\ddagger$ \\
Perplexity & No page fetched \\
Copilot    & No page fetched \\
\bottomrule
\end{tabular}

\smallskip
\raggedright\footnotesize
$^\dagger$Claude-User and ChatGPT-User had already visited this origin during API runs. Claude had requested the file 51\,min earlier, but we cannot attribute that request to its app trial.
$^\ddagger$An untagged \texttt{facebookexternalhit} request preceded Meta AI's page request, but its address differed from the page fetcher's.
\end{table}

\subsection{\texttt{robots.txt} Retention and Renewal}
\label{app:robots-reuse}

Among the user-triggered fetchers, only Anthropic requests \texttt{robots.txt} (\autoref{tab:fetcher-id}), so we ask how long it keeps the file and when it fetches it again.
To measure the retention, we asked Anthropic a question about a new page every 10 minutes and watched when it requested \texttt{robots.txt}.
In both of our runs it requested the \texttt{robots.txt} again 60 to 70 minutes after the initial request.
Anthropic therefore keeps its copy for about an hour, and a rule added to \texttt{robots.txt} may take up to about an hour to be seen.

\subsection{Public XML-Style Files}
\label{app:public-xml}

\begin{figure}[t]
  \centering
  \includegraphics[width=\linewidth]{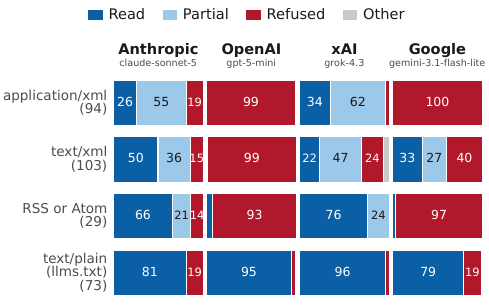}
  \caption{Can each service read public files of each declared type? Counts are in brackets.}
  \label{fig:public-xml-file-test}
\end{figure}

The public file test (\autoref{app:public-files}) asked each service to quote the first three lines of each public file (\autoref{app:prompts-public}).
\autoref{fig:public-xml-file-test} splits the answers for each service by the type the file declares, into four outcomes: the lines were quoted, \emph{partial} (only the XML header, or the file flattened into one line), refused, or \emph{other} (failed or inconclusive).
The findings by type are in \autoref{sec:content-type-filtering}.

\para{Why Google refused.}
Google read 34\% of the public files, mostly in full (\autoref{fig:public-xml-file-test}); \autoref{tab:gemini-public} shows why it did not read the rest.
It failed on 51\% without giving a reason, and did not finish the other 15\% because of capacity or a timeout.
Its failed answers only say that it cannot read the document, and none names a cause.
When Google failed, the model received only about 200 tokens, the size of an error message, so it never saw the file.

\begin{table}[t]
\centering
\small
\caption{Why Google and OpenAI did not read the public files.
Both columns are classified by the wording of the answer; Google's fetch status gives no cause either.}
\label{tab:gemini-public}
\begin{tabular}{@{}lcc@{}}
\toprule
Outcome & Google & OpenAI \\
\midrule
Refused, quoting its own fetch error & 0 & 192 \\
Refused, reporting the site's error (403, 404) & 0 & 16 \\
Refused, no reason given & 151 & 17 \\
\midrule
Quote limit (recitation) & 0 & 1 \\
Busy on both tries & 44 & 0 \\
Timed out or unclear & 2 & 1 \\
\midrule
Not read, total & 197 & 227 \\
\bottomrule
\end{tabular}
\end{table}

\subsection{Why Google's URL-Context Tool Failed on Our Origins}
\label{app:gemini-origins}
Google's \texttt{url\_context} tool is how the Gemini API reads a page from a URL.
We leave it out of the controlled-origin tests, because its answers did not show whether a page had been read.
As a control, we also asked Google to read \texttt{example.com}, a well-known page that any working fetcher can read.

\para{Two models, two problems.}
With \texttt{gemini-3.1-flash-lite}, Google's tool downloaded the whole page from our origin in nine calls, yet reported \texttt{ERROR} and the model said it could not access the page.
The \texttt{example.com} control also reported \texttt{ERROR} in every session, so it could not show that the tool worked.

\texttt{gemini-2.5-flash} behaved differently: when it failed, no request reached our origin.
But its results were inconsistent, although we changed nothing on our side.
It failed on Origin A for weeks before it read the page once, and on Origin B for several days before it read it.
A missing fetch therefore could not separate caching from Google's own failure.
The model is also deprecated: from 1 December 2026, Google routes requests to it to a different model~\cite{google2026deprecations}.
On the free tier, it allowed only 20 requests per day.

\para{Capacity.}
Google also answered that it was busy in a few calls.
All but one call succeeded on the retry.

\parai{Reuse.}
We could run one check, on \texttt{gemini-2.5-flash}, once it began to answer.\footnote{The free tier's daily limit allowed only a few such tests.}
We asked Gemini about the same page again from the same caller, 5 and 15 minutes after the first question.
Our origin saw a new download from Google's fetcher every time, so Gemini kept no copy.

\subsection{Token Cost of Failed Fetches}
\label{app:fault-tokens}
\begin{figure}[t]
  \centering
  \includegraphics[width=\columnwidth]{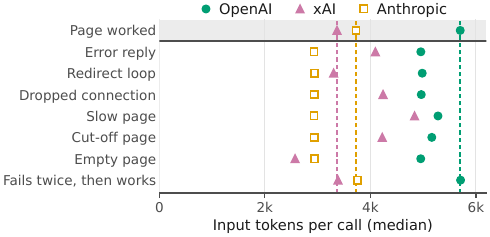}
  \caption{Input tokens per call under each fault.
  The line marks a successful fetch.
  Both origins pooled.}
  \label{fig:fault-token-cost}
\end{figure}

A failed fetch still costs input tokens, and we ask how much compared with a successful one.
\autoref{fig:fault-token-cost} shows the median input tokens per call under each fault, with both origins pooled.
The line marks a successful fetch of the same page.

On average, a failed fetch costs about as much as a successful one.
Anthropic and OpenAI are somewhat below the line.
xAI varies widely from fault to fault, so we see no pattern for it.

\section{Additional Caching and Reuse Results}
\label{app:caching}
This appendix supports \autoref{sec:cache}: how pages are matched across URLs (\autoref{app:url-equivalence}), shared across callers (\autoref{app:personalised-links}, \autoref{app:app-sharing}), and reused under cache headers (\autoref{app:cache-settings}), and how long they are kept (\autoref{app:renewal-test}, \autoref{app:public-staleness}, \autoref{app:endpoint-staleness}, \autoref{app:old-after-fetch}).

\subsection{URL Equivalence}
\label{app:url-equivalence}

This appendix asks which URLs a service treats as the same page.
If it treats two URLs as one, it can serve both from one cached copy (\autoref{sec:cache}).

For each change in \autoref{tab:url-equivalence}, we asked about a tagged URL, then about the changed URL about 90\,s later, in three trials with new tags.
If a new request reached our origin, the service had fetched the changed URL separately; if not, and the answer still held the page's content, it had reused the first copy.

\begin{table*}[t]
\centering
\small
\caption{Which URL changes each service treats as the same page?
$=$: reused the first copy; $\neq$: fetched anew; E: failed before retrieval.
Each result agreed across three trials.
The groups follow RFC~3986~\cite[§6]{rfc3986}, which does not prescribe a cache action.}
\label{tab:url-equivalence}
\begin{tabular}{@{}lllccc@{}}
\toprule
Change & Example edit & & OpenAI & Anthropic & xAI \\
\midrule
\multicolumn{6}{@{}l}{\emph{RFC~3986 leaves equivalence to the application}} \\
Add query item          & \texttt{\&x=1}                         & & $\neq$ & $\neq$ & $\neq$ \\
Remove whole query      & \texttt{?probe=t\&k=a} $\to$ none       & & $\neq$ & $\neq$ & $\neq$ \\
Reorder query items     & \texttt{a=1\&b=2} $\to$ \texttt{b=2\&a=1} & & $=$ & $=$ & $=$ \\
Add tracking item       & \texttt{\&utm\_source=email}            & & $\neq$ & $=$ & $\neq$ \\
Repeat item name        & \texttt{\&k=b}                          & & $\neq$ & $=$ & $\neq$ \\
Add empty item          & \texttt{\&z=}                           & & $\neq$ & $=$ & $\neq$ \\
Add trailing slash      & \texttt{/p} $\to$ \texttt{/p/}          & & $\neq$ & $\neq$ & $=$ \\
\midrule
\multicolumn{6}{@{}l}{\emph{RFC~3986: same retrieval target}} \\
Add fragment            & \texttt{\#entries}                     & & $=$ & $=$ & $=$ \\
Write default port      & \texttt{:443}                           & & $=$ & E & $=$ \\
Percent-encode letter   & \texttt{a} $\to$ \texttt{\%61}          & & $=$ & $=$ & $=$ \\
Add dot segment         & \texttt{/p} $\to$ \texttt{/./p}         & & $=$ & $=$ & $=$ \\
Capitalize hostname     & \texttt{h.test} $\to$ \texttt{H.TEST}   & & E & $=$ & $=$ \\
\midrule
\multicolumn{6}{@{}l}{\emph{RFC~3986: path is case-sensitive}} \\
Capitalize path         & \texttt{/p} $\to$ \texttt{/P}           & & $\neq$ & $\neq$ & $\neq$ \\
\bottomrule
\end{tabular}
\end{table*}

The services agree on some changes: a fragment, a percent-encoded letter, a dot segment, and a reordered query string all reuse the first copy at all three.
They differ where URL syntax leaves the decision to the application.
Anthropic reuses the copy when a tracking item, a repeated item name, or an empty item is added; xAI does so when a trailing slash is added; OpenAI does so in none of these cases.
We see no common rule behind these differences.

Two spellings failed before retrieval.
For the explicit \texttt{:443} port, Anthropic's fetch tool returned \texttt{url\_not\_in\_prior\_context} in all three trials, although the model passed the URL to the tool unchanged, port included.
The refusal therefore comes from the tool's own check, not from the model.
OpenAI reported that it could not open the capitalized hostname in all three trials; one answer quoted the hostname in lowercase.
These failures show neither reuse nor a separate copy.

No service reuses a copy when a query item is added or the whole query is removed.
So a random tag in the query string (\texttt{?probe=...}) makes a new page for all three services.
Our experiments therefore tag every trial's URL, so that the same page can be asked about again and again without earlier answers interfering (\autoref{sec:setup-probes}).

\subsection{Personalised Links and Cross-Account Leakage}
\label{app:personalised-links}
This appendix asks whether a page personalised by one account's link can reach another account, since the services share pages across accounts (\autoref{sec:content-reuse-and-sharing}).
The page's content depends on one query item in its URL, such as \texttt{?mc\_eid=user-1}.
Account A asked about the page with its own link.
Five minutes later, account B asked about it with a different link (\texttt{?mc\_eid=user-2}), and then with the item removed.
B receiving A's content without a new download would be a leak.

No leak occurred (\autoref{tab:personalised-links}).
OpenAI and xAI kept every item in the URL, so B's questions caused new downloads and B received its own content.
Anthropic dropped \texttt{utm\_content}, \texttt{\_hsenc}, and \texttt{fbclid} before downloading, so the personalised page was never fetched for anyone and every account got one generic copy; it kept \texttt{mc\_eid} and \texttt{uid} as separate URLs.
This matches the tracking-item row of \autoref{tab:url-equivalence}.
We tested query items only, through the APIs; cookies, headers, and the consumer apps were not tested.

\begin{table}[t]
\centering
\small
\setlength{\tabcolsep}{4pt}
\caption{What each fetcher did with a query item?
\textbf{Dropped}: removed before downloading, so every account got the same generic copy.
Kept: left in the URL, so a different value caused a new download.
No leak occurred.}
\label{tab:personalised-links}
\begin{tabular}{@{}llll@{}}
\toprule
Query item & Anthropic & OpenAI & xAI \\
\midrule
\texttt{utm\_content} & \textbf{Dropped} & Kept & Kept \\
\texttt{mc\_eid}      & Kept & Kept & Kept \\
\texttt{\_hsenc}      & \textbf{Dropped} & Kept & Kept \\
\texttt{fbclid}       & \textbf{Dropped} & Kept & Kept \\
\texttt{uid}          & Kept & Kept & Kept \\
\bottomrule
\end{tabular}
\end{table}

\subsection{Consumer-App Content Reuse}
\label{app:app-sharing}

We repeated the sharing test in eight consumer apps, using two separate free logins, A and B, per app.
Login A asked question A1 about page A, and about 7 minutes later asked question A2 about it in a new chat (reuse within a login).
About 14 minutes after A1, login B asked question B1 about page A (sharing across logins), and then question B2 about a fresh page B.
Each question asks for a different code from its page, and a correct code with no request to our origin before the answer counts as reuse or sharing.

\begin{table}[t]
\centering
\small
\caption{Content reuse in consumer apps: was the page answered with no new request to our origin? Yes: in all three runs; No: in none; otherwise the number of runs.}
\label{tab:app-sharing}
\begin{tabular}{@{}lcc@{}}
\toprule
 & \multicolumn{2}{c}{Reused?} \\
\cmidrule(l){2-3}
App & New chat, same login & Different login \\
\midrule
Claude   & Yes    & Yes    \\
ChatGPT  & Yes    & Yes    \\
Gemini   & Yes    & No     \\
Grok     & 1 of 3 & 2 of 3 \\
DeepSeek & No     & No     \\
Mistral  & Yes    & Yes    \\
Qwen     & No     & No     \\
Meta AI  & No     & No     \\
\bottomrule
\end{tabular}
\end{table}

Three apps shared the page across logins (\autoref{tab:app-sharing}): Claude, ChatGPT, and Mistral, every time, and they also reused it within a login.
Gemini reused it only within a login.
DeepSeek, Qwen, and Meta AI fetched it again every time.
Grok was inconsistent: it shared the page in two of three runs and reused it within a login in one.

\subsection{Cache-Header Experiment: All Settings}
\label{app:cache-settings}
This appendix lists every \texttt{Cache-Control} setting we tested and what each service did with it (\autoref{sec:cache-directives}).
We served 19 HTML pages of 2\,KB that differ only in their codes and their \texttt{Cache-Control} setting, and asked about each again at intervals that double from 2.5 to 80 minutes.

\autoref{tab:cache-directives} shows, for each setting, what RFC~9111 says a shared cache should do and whether each service reused the page.
OpenAI reused every page for about 30\,min and then fetched it again, xAI reused every page whatever its setting, and Anthropic did the same except under \texttt{no-store} from 17 October.

\begin{table*}[t]
\centering
\small
\setlength{\tabcolsep}{5pt}
\renewcommand{\arraystretch}{1.15}
\caption{Cache-header experiment: was a page answered from a stored copy, with no origin request, when asked again up to 80\,min later? One page per setting, service, and run. $\to$: result changed on 17 October. \texttt{Last-Modified}: old = a date in the past; recent = the origin's start time; absent = no \texttt{Last-Modified} or \texttt{ETag}. Where a row lists several, the result was the same for each variation.}
\label{tab:cache-directives}
\begin{tabular}{@{}>{\raggedright\arraybackslash}p{4.9cm}>{\raggedright\arraybackslash}p{2.5cm}>{\raggedright\arraybackslash}p{4.8cm}ccc@{}}
\toprule
 & & & \multicolumn{3}{c}{Reused?} \\
\cmidrule(l){4-6}
\texttt{Cache-Control} & \texttt{Last-Modified} & RFC~9111 shared-cache rule & OpenAI & Anthropic & xAI \\
\midrule
\multicolumn{6}{@{}l}{\emph{Reuse forbidden without contacting the origin}} \\
\texttt{no-store} & old / absent & Do not store or reuse & Yes & Yes $\to$ No & Yes \\
\texttt{no-cache, no-store, max-age=0, must-revalidate} & old / absent & Do not store or reuse & Yes & Yes $\to$ No & Yes \\
\texttt{private} & old / absent & Shared cache must not store & Yes & Yes & Yes \\
\texttt{no-cache} & old / absent & Validate before every reuse & Yes & Yes & Yes \\
\texttt{max-age=0} & old / absent & Stale immediately; validate or fetch & Yes & Yes & Yes \\
Past \texttt{Expires} & old / absent & Stale immediately; validate or fetch & Yes & Yes & Yes \\
\midrule
\multicolumn{6}{@{}l}{\emph{Reuse allowed for a stated time}} \\
\texttt{max-age=60} & old & Fresh for one minute & Yes & Yes & Yes \\
\texttt{max-age=60, must-revalidate} & old & Validate once stale after one minute & Yes & Yes & Yes \\
\texttt{max-age=3600, s-maxage=60} & old & Shared cache: validate after one minute & Yes & Yes & Yes \\
\texttt{max-age=3600} & old & May reuse while fresh for one hour & Yes & Yes & Yes \\
\midrule
\multicolumn{6}{@{}l}{\emph{No reuse directive}} \\
none & old / recent / absent & Heuristic freshness may apply & Yes & Yes & Yes \\
\bottomrule
\end{tabular}
\end{table*}

\subsection{Renewal Test}
\label{app:renewal-test}
This appendix asks whether a page that is asked about often keeps its stored copy longer than one that is not.
For OpenAI and xAI it does not: asking again did not extend the copy, so how stale an answer is does not depend on how busy the page is.

We used three pages.
The \emph{queried} page is asked about three times before its copy should expire (at about 7, 14, and 21 minutes for OpenAI, and 6, 12, and 18 hours for xAI), and the \emph{untouched} page only at the start and at the end.
The \emph{late} page is first asked about shortly before the end, as a quick check that the service still reuses copies at that moment.
At the final check, the origin log shows which pages the service downloaded again.

If questions extended a copy, the queried page would still be served from its copy while the untouched page was downloaded again.
Instead, in all three runs both were downloaded again, for OpenAI at 36--43 minutes and for xAI at 32--33 hours, while the late page was still served from its copy.

We did not run the test for Anthropic, because it kept every HTML copy for over three days (\autoref{sec:retention-and-renewal}), too long for this test.

\subsection{Staleness on Public Pages}
\label{app:public-staleness}

We also tried to measure staleness on real pages.
It works for Anthropic, whose reply says when it fetched each page, but not for the other services: nothing in their answers shows what was fetched or when, so we could not tell a stale answer from a wrong one with confidence.
This is why the main staleness result uses public endpoints with known values (\autoref{app:endpoint-staleness}).
We report the attempt because it shows how hard freshness is to judge from outside.

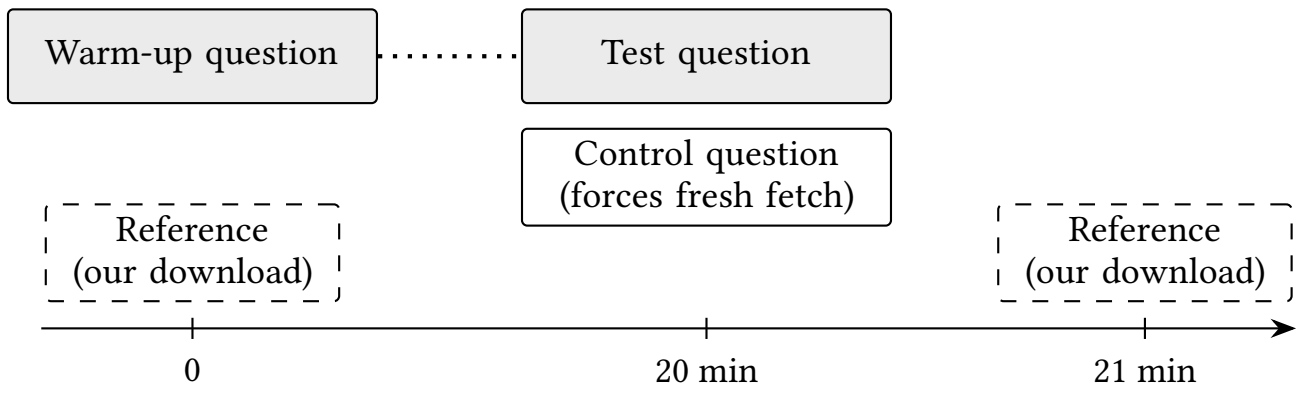
\begin{figure}[t]
\centering
\begin{tikzpicture}[font=\scriptsize,>=Stealth,
  q/.style={draw,rounded corners=1pt,fill=black!8,minimum height=0.62cm,align=center,inner sep=2pt,text width=2.3cm},
  ctl/.style={q,fill=white},
  ref/.style={draw,dashed,rounded corners=1pt,minimum height=0.5cm,align=center,inner sep=2pt,text width=1.8cm}]
\draw[->] (0,0) -- (8.3,0);
\foreach \x/\l in {1.0/0,4.4/20 min,7.3/21 min}{\draw (\x,0.07)--(\x,-0.07) node[below]{\l};}
\node[q]   (warm) at (1.0,1.8) {Warm-up question};
\node[q]   (test) at (4.4,1.8) {Test question};
\node[ctl] (ctrl) at (4.4,1) {Control question\\(forces fresh fetch)};
\node[ref] at (1.0,0.5) {Reference\\(our download)};
\node[ref] at (7.3,0.5) {Reference\\(our download)};
\draw[dotted,thick] (warm.east) -- (test.west);
\end{tikzpicture}
\caption{One public-page trial. The warm-up question may leave a copy (dotted line) that the test question reuses. Dashed boxes are our own downloads, the reference for both answers.}
\label{fig:public-trial}
\end{figure}

\para{Design.}
We want to know whether a service answers from an old copy of a real page.
Public pages have no changing field that is easy to read and consistent across sites, so we first locate one: an anchor, a line just above the part that changes often (\autoref{app:public-pilot}).
Each trial then asks the service to quote the words after the anchor, twice, and grades both answers against our own downloads of the page (\autoref{fig:public-trial}).

The \emph{test} question uses the page's normal URL, 20 minutes after a warm-up question that gave the service a copy of the page.
If the page has changed since, a service that reuses the copy gives an old answer.
The \emph{control} question is asked at the same time, at a URL with a random tag, which makes the services fetch again (\autoref{app:url-equivalence}).
It shows what a current answer looks like, and how often the service misses even when the page is fresh.
A download is our own request for the page, and a question goes to the service; the downloads, taken around the questions, show what the page said.
We run this for each of 90 pages (\autoref{sec:setup}).

\para{Judging an answer.}
We label an answer \emph{stale} in one case only: the control answer matches the page and the test answer does not.
An answer matches if it shares at least 5 consecutive words (10 characters for Japanese and Chinese) with the first 60 words (180 characters) after the anchor in our download.
Matches are reliable: no answer matched our download of a different page.
The stale count is therefore a lower bound.
Most pairs fall in the other cases:
\begin{smartlist}
\item{The model cannot read the page.}
In 13\% (OpenAI) to 46\% (xAI) of answers said they could not read it; this affects every question, whether warm-up, test, or control.
\item{The model gives the wrong text.}
Even after reading the page, it may quote the wrong lines, paraphrase, or quote from elsewhere on the page.
\item{The page changes between the control question and our download.}
If a page changes often, our reference download may catch a different version than the control question saw, so we cannot check the control answer against it.
\end{smartlist}
As a trust check, we also count \emph{control misses}: pairs where the test answer matches the reference download but the control answer does not.
The control should always match, so these show how often an answer misses for reasons other than staleness; we call staleness only where stale answers clearly outnumber the \emph{control misses}.

\para{Anthropic.}
Anthropic is the one service that lets us check our method, because its replies give the time it fetched each page and the text it received.
We first check our baseline.
Every copy behind a control answer was fetched fresh, so the random tag does force a new fetch.
Our stale label is weaker: it has a precision of 76\% and a recall of 50\%, so it under-counts.
We then ask how old the copies behind the test answers are.
For 65\% of them, Anthropic had fetched the page more than an hour before our question, and these pages change often.
\footnote{We also checked that old copies really differ from the live page.
We measure how much of each copy's text appears in our own download of the page, taken at question time.
The older the copy, the lower this overlap: the median is 64--69\% for copies under an hour old, close to the 70\% of control answers, and 45--48\% for copies between one hour and seven days old.}
So Anthropic's answers are stale, in line with the counter endpoints (\autoref{app:endpoint-staleness}).
Our label agrees: 20 pages were stale against 3 control misses (exact binomial $p=0.0005$).
For 21\% of the pages Anthropic returned no copy, mostly because it refuses the site (\texttt{url\_not\_allowed}), which is the cannot-read problem above.

\para{The other services.}
For the other three, the counts in \autoref{tab:public-staleness} do not settle the question, because stale answers do not clearly outnumber control misses.
xAI leans stale (12 against 4, $p=0.077$), while OpenAI has 10 against 9 and Google 17 against 13.
Many pairs give no comparison at all, because neither answer matches the page, for the reasons listed above.

\begin{table}[t]
\centering
\footnotesize
\setlength{\tabcolsep}{3pt}
\caption{Test and control answers for 90 pages per service, judged against our reference download. $p$: exact binomial test of stale answers against control misses.}
\label{tab:public-staleness}
\begin{tabular}{@{}lccrrrr@{}}
\toprule
 & \multicolumn{2}{c}{Match} & & & & \\
\cmidrule(lr){2-3}
Outcome & Test & Ctrl & Anthr. & OpenAI & xAI & Google \\
\midrule
Stale        & No  & Yes & 20 & 10 & 12 & 17 \\
Control miss & Yes & No  &  3 &  9 &  4 & 13 \\
Both match   & Yes & Yes & 21 & 43 & 23 &  7 \\
Neither      & No  & No  & 44 & 28 & 51 & 48 \\
Unusable     & \multicolumn{2}{c}{--} & 2 & 0 & 0 & 5 \\
\midrule
$p$ & & & 0.0005 & 1.0 & 0.077 & 0.58 \\
\bottomrule
\end{tabular}
\end{table}

\para{Why we cannot see more.}
Apart from Anthropic, the services tell us little about what they fetched.
Anthropic reports the fetch time and an error code.
Google reports only a status, success or error, with no reason; its \textsc{error} roughly matches its ``cannot read'' answers (20\% against 23\%).
OpenAI and xAI, which we call through OpenRouter, report nothing but a citation of the page.
They attach it even when they then say they could not read the page: xAI cites the page in nearly every answer and still says it cannot read it in 46\%.
A citation is therefore not a usable signal of failure: it appears whether or not the service got the page.

\para{A check we could not use.}
We also tried to judge each answer against the other services' fresh answers, instead of our own download.
This failed, because the services open their replies with similar stock phrases and paraphrase in similar ways, so their answers overlap even when they are about different pages.
This shared wording drowns out any real agreement.

\para{What this shows.}
On real pages, without a fetch time or a reliable error report, an outside observer cannot tell a stale answer from a wrong one with confidence.
Only Anthropic gives us that, and even there our label under-counts.

\subsection{Public-Endpoint Staleness: All Endpoints}
\label{app:endpoint-staleness}

This appendix extends the endpoint results in \autoref{sec:retention-and-renewal}.
Unlike public pages (\autoref{app:public-staleness}), whose text changes in unpredictable ways, these endpoints return a number that only increases, so every answer can be checked and dated against our reference log.
Below, we describe how we measured staleness, show the result on every endpoint, and explain the cells that do not follow the usual pattern.

We asked each service for the current value of eight public endpoints every 45 minutes for about three days.
To date the answers, we polled the endpoints ourselves every five minutes, so staleness below five minutes cannot be resolved.
For OpenAI, xAI, and Google, staleness is how long ago the value they gave stopped being the live value; Anthropic reports when it fetched each page (\texttt{retrieved\_at}), so we use the time since that fetch.
Two cells are excluded.
Anthropic refused StackOverflow in nearly every round: its fetch tool reported the URL as not allowed (\texttt{url\_not\_allowed}), a policy block and not a failed fetch~\cite{anthropic2025webfetch}.
Google's USGS answers are excluded because the value matches no value the feed ever had: its fetch reported an error in every round, and the timestamps it gave match the time of the question except for the year, which is two to four years in the past.

\begin{figure*}[t]
  \centering
  \includegraphics[width=\textwidth]{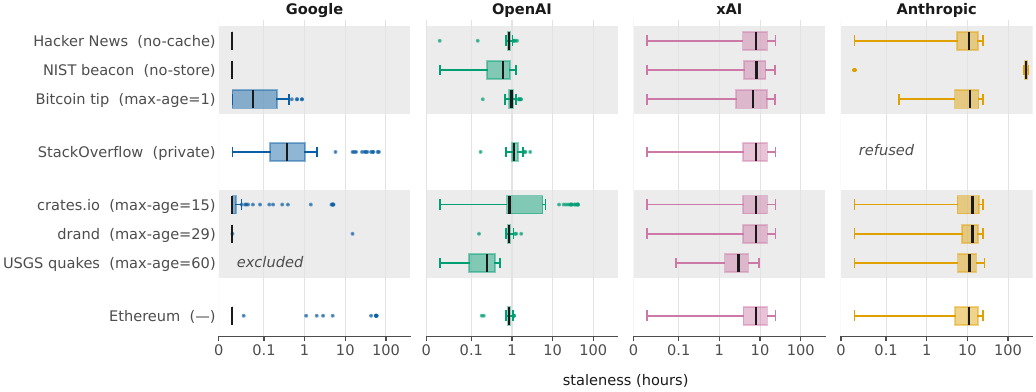}
  \caption{How far behind the live value each service's answer is, on eight public endpoints.
  Shading groups endpoints by their \texttt{Cache-Control} header.
  Boxes show quartiles, lines the median, and dots outliers.
  For most services the median does not change with the header.
  Anthropic refused StackOverflow, and Google's USGS answers are excluded.}
  \label{fig:endpoint-staleness-all}
\end{figure*}

As shown in \autoref{fig:endpoint-staleness-all}, on most endpoints, median staleness is near zero for Google, about an hour for OpenAI, about eight hours for xAI, and about eleven for Anthropic.
The exceptions are of two kinds, by what the fetch did:

\para{An old value although the fetch succeeded.}
OpenAI on crates.io returned the same value over many rounds, far beyond the hour its answers usually lag, which gives it a tail of about 40 hours.
Anthropic on the NIST beacon is the stronger case: in most rounds it answered from a copy retrieved on 24 October, more than ten days before the rounds we compare against, although the page sends \texttt{no-store}.
In the other rounds its fetch returned the live value or refused.\footnote{This only partly matches our controlled tests, where Anthropic fetched \texttt{no-store} pages fresh every time (\autoref{tab:anthropic-nostore}).}
The old copy comes from the fetch tool's own cache, which Anthropic documents and says may return content that is not the latest~\cite{anthropic2025webfetch}.
As a quick check, we also confirmed that Anthropic can retrieve the fresh page.\footnote{With the latest version of the tool (\texttt{web\_fetch\_20260309} or later) and \texttt{use\_cache} set to \texttt{false}; all other runs use the basic version.}
We cannot vary the server-side setup, because NIST is not ours.

\para{An old value after a failed fetch.}
Google's tails on StackOverflow and Ethereum height (up to about 60 hours) come from rounds where its fetch reported an error and it returned an outdated value.
After a successful fetch, its median answer was within minutes of the live value; after a failed fetch, it was more than a day behind.

\begin{table}[t]
\centering
\small
\caption{The public endpoints we probed: the \texttt{Cache-Control} header each sends, the host or CDN behind it (inferred from response headers), and how often its value grows by one step. We estimate the last from the growth between our polls and round it. USGS reports a timestamp, so its value moves continuously.}
\label{tab:anchor-staleness}
\begin{tabular}{llll}
\toprule
Endpoint & \texttt{Cache-Control} & Host/CDN & Grows every \\
\midrule
Hacker News     & \texttt{no-cache}    & Firebase   & 7\,s \\
NIST beacon     & \texttt{no-store}    & Cloudflare & 1\,min \\
Bitcoin tip     & \texttt{max-age=1}   & nginx      & 5\,min \\
StackOverflow   & \texttt{private}     & Cloudflare & 2\,min \\
crates.io       & \texttt{max-age=15}  & Heroku     & 2\,min \\
drand           & \texttt{max-age=29}  & nginx      & 30\,s \\
USGS quakes     & \texttt{max-age=60}  & CloudFront & continuous \\
Ethereum height & ---                  & Cloudflare & 12\,s \\
\bottomrule
\end{tabular}
\end{table}

\subsection{Old Answers After a New Fetch}
\label{app:old-after-fetch}

\para{OpenAI often answers from the old version after fetching the new one.}
An \emph{outdated answer} quotes the old version of a page although the service fetched the new version during the same question; an \emph{updated answer} quotes the new version.
In the page-update and error-fix tests (\autoref{fig:timelines}), OpenAI gave an outdated answer in about 60\% of the follow-up questions.
Anthropic never did, and xAI did in 2\% of the error-fix follow-ups.

\para{Outdated answers are cheaper and faster, as if no new page was read.}
Compared with updated answers, outdated answers made fewer search calls, used about a third fewer input tokens, and took less time.
In the error-fix test, outdated answers still reported the original error and never cited the recovered page.
In the one case where the answer leaked the tool's internal name for the page it opened (\texttt{turn\dots\allowbreak view\dots}), it quoted the code of the old version, although our origin had served the new one in the same question.

\para{A possible explanation: the tool reads a stored copy, and the fetch refreshes it.}
We cannot see inside OpenAI's tool, because its reasoning is encrypted and its browsing steps are hidden.
One explanation fits what we saw: the tool answers from a copy it already holds, and the fetch we see only refreshes that copy for the next question.
Three observations fit it:
\begin{smartlist}
\item{Cost.}
Outdated answers cost less, as if no new page was read; an updated answer costs more, as if a second call read the refreshed copy and added its content to the input.
\item{Timing.}
Our origin received exactly one fetch per question, about three seconds in, whether the answer was outdated or updated, so the fetch does not decide the answer.
In the error-fix test, we kept asking after the fix, and for about 25 minutes OpenAI answered with the original error without sending any request to our origin.
\item{Resemblance.}
It looks like serving a stale copy while it is revalidated~\cite{rfc5861}, and OpenAI's open-source reference browser likewise keeps opened pages~\cite{openai2025gptoss}.
\end{smartlist}
Nothing says the deployed service works the same way.

If this is right, and the stored copy is shared across accounts (\autoref{sec:content-reuse-and-sharing}), a fetch triggered by one caller's question mainly benefits the next caller: the first is answered from the old copy, the later one from the refreshed copy.
For a publisher, a fetch in the origin log then does not show that the page reached the user who caused it, and a change may reach users one question late.
\section{Stakeholder Impact Rationale}
\label{app:stakeholder-rationale}

\autoref{tab:stakeholders} rates each finding from the viewpoint of four stakeholders; this appendix gives the reasoning behind every rating.
We judge each stakeholder by what it cares about.
A \emph{web publisher} cares about the load on its origin, its control over how its content is reused, and whether its pages are reachable.
An \emph{LLM vendor} cares about fetch cost; we do not rate vendors on answer quality.
An \emph{end user} cares about whether answers are correct, fresh, and transparent about their sources and cost.
\emph{CDN and infrastructure operators} care about cache efficiency and the traffic they must absorb.
A rating is negative, positive, or mixed in that sense, and a cell is left blank when the finding changes nothing for that stakeholder.
The ratings are our assessment of the measured behavior, not a measurement of harm.
\autoref{tab:stakeholders-rationale} has one row for every rated cell.

\begin{table*}[t]
\centering
\caption{Why each cell of \autoref{tab:stakeholders} has its rating.
  Citations in parentheses are user-reported issues observed in the wild, not prior measurement studies.}
\label{tab:stakeholders-rationale}
\footnotesize
\setlength{\tabcolsep}{4pt}
\renewcommand{\arraystretch}{1.1}
\begin{tabular}{@{}L{3.0cm} L{1.9cm} G{0.7cm} L{11.3cm}@{}}
\toprule
\textbf{Finding} & \textbf{Stakeholder} & & \textbf{Reasoning} \\
\midrule

\texttt{robots.txt} skipped by OpenAI and xAI (\autoref{sec:fetcher-id})
  \Pub{\cellbad}{\texttt{robots.txt} is where a site lists what bots may read. OpenAI and xAI never request it, so a site cannot limit them this way.}
\cmidrule{1-4}

\texttt{Retry-After} ignored by OpenAI and xAI (\autoref{sec:failure-retries})
  \Pub{\cellbad}{A server can send \texttt{Retry-After} to ask clients to wait before trying again. OpenAI and xAI ignored it, so an overloaded site gets no relief.}
  \CDN{\cellbad}{The CDN in front of the site cannot slow these fetchers down this way, so it has to absorb their extra requests itself.}
\cmidrule{1-4}

xAI sends parallel requests (\autoref{sec:failure-retries})
  \Pub{\cellbad}{One user question makes xAI send several requests at the same time, which multiplies the load on the site.}
  \CDN{\cellbad}{The requests come from many different addresses and look like browsers, so the CDN can hardly tell them from visitors or limit them~\cite{wikimedia2025crawlers,zhang2025webcache}.}
\cmidrule{1-4}

Anthropic stops after one failure (\autoref{sec:failure-retries})
  \Pub{\cellmixed}{Anthropic sends one request and stops, which adds little load, but a page that would have worked on a retry loses the chance to be read.}
  \User{\cellbad}{After a temporary error Anthropic does not retry, while OpenAI and xAI do, so its answer can miss a page that was available (e.g.,~\cite{young2025browse500}).}
\cmidrule{1-4}

OpenAI discards some formats (\autoref{sec:content-type-filtering})
  \Pub{\cellbad}{OpenAI's requests accept any format, yet it drops pages served as Markdown, XML, CSV, or DOCX after fetching them.
  The publisher sends the page, and it never informs an answer.}
  \User{\cellbad}{The answer rests on fewer sources than the user would expect, and nothing tells them that a page was left out.}

Script-only content not read (\autoref{sec:javascript})
  \Pub{\cellbad}{No service reads text that a page adds with JavaScript or puts in a script block (for example JSON-LD, used by about 40\% of popular homepages).}
  \User{\cellbad}{The answer lacks content that a person browsing the page would see.}
\cmidrule{1-4}

xAI reads only the first 0.5\,MB (\autoref{sec:info-access-accross-doc-size})
  \Pub{\cellbad}{xAI requests the whole page but reads only its first 0.5\,MB, so the rest is sent and then dropped from the answer.}
  \Vend{\cellgood}{Capping what is read keeps the model's input, and so the cost of each answer, bounded.}
  \User{\cellbad}{The answer omits the rest of a long page without saying so (e.g.,~\cite{bethke2026firstkilobyte}).}
\cmidrule{1-4}

Input cost grows with page size (\autoref{sec:info-access-accross-doc-size}, \autoref{sec:post-op-cost-amp})
  \User{\cellbad}{With Anthropic, one question about a 2\,MB page can use about 760\,K tokens, and a failed fetch costs nearly as much as a successful one. The caller cannot predict this before asking.}
\cmidrule{1-4}

Pages shared across accounts (\autoref{sec:content-reuse-and-sharing})
  \Pub{\cellmixed}{A page fetched for one person's question is reused for others, so the site receives fewer requests. But when a site is paid per fetch, one fetch can serve many answers~\cite{teixeira2026payperuse}; whether the site is credited for each use depends on the vendor's attribution, not on the sharing itself.}
  \Vend{\cellgood}{Fewer fetches means a lower fetch cost.}
  \User{\cellmixed}{Your answer may rest on a page fetched for someone else. We found no leak of personalised links, but tested only links personalised in the URL, and the services do not say how widely pages are shared (e.g.,~\cite{mudgal2026memleak}).}
  \CDN{\cellgood}{Fewer repeated requests for the same page reach the CDN.}
\cmidrule{1-4}

Site's cache rules ignored (\autoref{sec:cache-directives}, \autoref{sec:retention-and-renewal})
  \Pub{\cellbad}{\texttt{Cache-Control} is how a site says how long its page may be reused, and the services mostly ignore it. An update reaches answers after 30 minutes to a day or more, and the site cannot change that.}
  \Vend{\cellgood}{Ignoring it means fewer fetches than obeying short limits would cost.}
  \User{\cellbad}{Answers can come from an old copy, and the user is not told its age and cannot ask for a fresh fetch (e.g.,~\cite{ivan2025stalesearch}).}
  \CDN{\cellbad}{A CDN relies on these headers to know when a copy is out of date; the AI services bypass them and use their own copy lifetimes~\cite{teixeira2026payperuse}.}
\cmidrule{1-4}

Failed pages stay failed (\autoref{sec:neg-cache})
  \Pub{\cellbad}{After a page fails and is then fixed, services keep answering as if it were broken, so a working page stays out of answers.}
  \User{\cellbad}{The answer says the page is down when it works.}
\cmidrule{1-4}

OpenAI answers from an old copy (\autoref{app:old-after-fetch})
  \Pub{\cellbad}{OpenAI often fetches the new version of a page and still answers from the old one, so a fetch in the site's log does not mean the user saw the page.}
  \User{\cellbad}{The answer is old even though the page had just been refreshed.}
\bottomrule
\end{tabular}
\end{table*}
\section{Open Science}
\label{sec:open-science}

We release the artifacts of this study in public repositories: code at \codeurl{} and data at \dataurl{}.
\begin{smartlist}
\item{Test pages and probe code.} The synthetic pages (\autoref{app:test-materials}), the prompts, and the code that sends each question and checks the answer.
\item{Origin logs.} The request records that carry our URL tags. Each source address is replaced by a random ID, and we keep the network number, owner, and country (\autoref{app:source-networks}).
\item{Vendor answers.} Every answer used in the results, with each question and its time, and Anthropic's reported fetch time where it gives one.
\item{Public-site data.} For the public-page tests, our reference downloads' values and times and each vendor's answer; for the public counters, each value and its time.
\end{smartlist}
Each page's code is released in a table, so that readers can re-check whether an answer quotes the page; the secret that generated the codes is not released.
We do not release API keys or account details.

\end{document}